\documentclass{emulateapj}
\usepackage{color} 

\usepackage{epsfig}
\usepackage{subfigure}
\usepackage{graphicx}
\usepackage{amsmath}
\usepackage{natbib}
\newcommand{\pa}{\partial}
\newcommand{\mb}{\boldsymbol}
\newcommand{\mc}{\mathcal}

\shorttitle{MRI Driven Accretion in PPDs}
\shortauthors{X.-N. Bai}

\begin{document}


\title{Magnetorotational Instability Driven Accretion in Protoplanetary Disks}


\author{Xue-Ning Bai}
\affil{Department of Astrophysical Sciences, Princeton University,
Princeton, NJ, 08544} \email{xbai@astro.princeton.edu}




\begin{abstract}
Non-ideal MHD effects play an important role in the gas dynamics in protoplanetary disks
(PPDs). This paper addresses its influence on the magnetorotational instability (MRI) and
angular momentum transport in PPDs using the most up-to-date results from numerical
simulations. We perform chemistry calculations using a complex reaction network with
standard prescriptions for X-ray and cosmic-ray ionizations. We first show that no matter
grains are included or not, the recombination time is at least one order of magnitude less
than the orbital time within $5$ disk scale heights, justifying the validity of local ionization
equilibrium and strong coupling limit in PPDs. The full conductivity tensor at different disk
radii and heights is evaluated, with the MRI active region determined by requiring that (1)
the Ohmic Elsasser number $\Lambda$ be greater than $1$; (2) the ratio of gas to magnetic
pressure $\beta$ be greater than $\beta_{\rm min}(Am)$ as identified in the recent study by
Bai \& Stone (2011), where $Am$ is the Elsasser number for ambipolar diffusion. With full
flexibility as to the magnetic field strength, we provide a general framework for estimating the
MRI-driven accretion rate $\dot{M}$ and the magnetic field strength in the MRI-active layer.
We find that the MRI-active layer always exists at any disk radius as long as the magnetic
field in PPDs is sufficiently weak. However, the optimistically predicted $\dot{M}$ in the inner
disk ($r=1-10$ AU) appears insufficient to account for the observed range of accretion rate
in PPDs (around $10^{-8}M_{\odot}$ yr$^{-1}$) even in the grain-free calculation, and the
presence of solar abundance sub-micron grains further reduces $\dot{M}$ by one to two
orders of magnitude. Moreover, we find that the predicted $\dot{M}$ increases with
radius in the inner disk where accretion is layered, which would lead to runaway mass
accumulation if disk accretion is solely driven by the MRI. Our results suggest that stronger
sources of ionization, and/or additional mechanisms such as magnetized wind are needed
to explain the observed accretion rates in PPDs. In contrast, our predicted $\dot{M}$
is on the order of $10^{-9}M_{\odot}$ yr$^{-1}$ in the outer disk, consistent with the observed
accretion rates in transitional disks.
\end{abstract}


\keywords{accretion, accretion disks --- instabilities --- magnetohydrodynamics (MHD) --- methods: numerical
--- protoplanetary disks --- turbulence}

\section{Introduction}\label{sec:intro}

Protoplanetary disks (PPDs) around pre-main-sequence stars form from the collapse
of protostellar cores as a result of angular momentum conservation \citep{Adams_etal87}.
With a typical lifetime of $1-10$ Myrs (e.g., \citealp{Hillenbrand_etal98,Sicilia_etal06}),
PPDs feed gas onto the central protostar, power an outflow and/or jet, and provide the raw
materials for the formation of planetary systems. The structure, evolution and dispersal of
PPDs are of crucial importance in understanding a wide range of physical problems
especially in the area of planet formation (see review by \citealp{Armitage11}). As more
and more exoplanets are discovered\footnote{see http://exoplanet.eu/}, together with the
advancement of planet formation theory (e.g., see the book by \citet{Armitage10}, and
also see \citet{ChiangYoudin10} for a review on planetesimal formation), understanding
the gas dynamics in more detail in the PPDs becomes essential.

One of the most important observational constraints relevant to the gas dynamics in PPDs
is that PPDs are actively accreting. The accretion signature comes from the UV excess
emission that veils the intrinsic photospheric spectrum of a YSO, which is interpreted as
coming from the standing accretion shock formed at the stellar surface
\citep{CalvetGullbring98,Gullbring_etal00}. More commonly, the accretion rate $\dot{M}$
can be inferred from emission line profiles, in particular the H$\alpha$ line, based on the
magnetospheric accretion model \citep{Muzerolle_etal98,Muzerolle_etal01}. The inferred
$\dot{M}$ for Classical T-Tauri stars is about $10^{-8\pm1} M_{\odot}$ yr$^{-1}$
\citep{Hartmann_etal98,Calvet_etal04,Sicilia_etal05}, and observations over a wide mass
range of protostars reveal a correlation between $\dot{M}$ and the protostellar
mass \citep{Muzerolle_etal05,Natta_etal06,HerczegHillenbrand08,Fang_etal09}.
Transitional disks, characterized by optically thin inner holes which may represent a later
evolutionary stage of PPDs \citep{Calvet_etal02,Calvet_etal05,Hughes_etal07,Espaillat_etal07},
are also observed to be actively accreting. Although transitional objects are much rarer, the
median of their accretion rate from currently available samples is a few times $10^{-9}
M_{\odot}$ yr$^{-1}$ \citep{Najita_etal07,Sicilia_etal10}. A key question here is, what drives
the rapid accretion in PPDs?

The answer may depend on the evolutionary stages of the PPDs, but it appears certain that
turbulence plays a crucial role, and evidence of turbulence in PPDs has been reported from
sub-millimeter interferometric observations \citep{Hughes_etal11}.
In early phases, the accretion is likely to be primarily driven by the gravitational instability
(GI). The GI leads to gravitoturbulence if the cooling rate is less than the orbital frequency,
and transports angular momentum outward via non-axisymmetric spiral waves very
efficiently \citep{Gammie01,Rice_etal05}. For typical accretion rate in PPDs, and given the
opacity dominated by dust grains, gravitoturbulence is likely to be present at intermediate
disk radii between a few tens and $\sim100$ AU
\citep{VorobyovBasu07,Rafikov09,Rice_etal10}. However, time-dependent calculations
of the PPD evolution indicate that after the envelope infall stops ($\lesssim1$ Myr), the
disk is generally not massive enough to sustain the GI \citep{Zhu_etal10b}. Other driving
mechanisms at the inner disk as well as beyond the infall (embedded) phase are clearly
needed.

The most important mechanism for driving accretion in non-self-gravitating thin disks
is believed to be the magnetorotational instability (MRI, \citealp{BH91}), which generates
turbulence and efficiently transports angular momentum radially outward
\citep{HGB95,SHGB96}. The MRI requires sufficient coupling between the gas and the
magnetic field. However, PPDs are only weakly ionized. The main ionization sources such
as cosmic rays and X-rays from the protostar only effectively ionize the surface layer of the
disk, making the surface layers ``active" to MRI driven turbulence, while the gas in the
midplane remains poorly coupled to the magnetic field and is termed `dead" \citep{Gammie96}.
The layered accretion scenario has been studied in great detail via both numerical
calculations in a fixed disk model with a chemical reaction network \citep{Sano_etal00,
Fromang_etal02,Semenov_etal04,IlgnerNelson06,BaiGoodman09,TurnerDrake09}, and
MHD simulations \citep{FlemingStone03,Turner_etal07,TurnerSano08,IlgnerNelson08,
OishiMacLow09}. These research works generally confirm that a ``dead zone" is
expected at the inner part of the PPDs (about $0.5-10$ AU), though its radial and vertical
extent depends on the ionization rate and the abundance of small (sub-micron) dust
grains.
 
Because of the low ionization level in PPDs, the gas dynamics is controlled by a
number of non-ideal MHD effects, including Ohmic resistivity, Hall effect and ambipolar
diffusion (see \citet{Balbus11} for a review). All calculations and simulations mentioned
above considered only the effect of Ohmic resistivity. However, Hall effect and ambipolar
diffusion (AD) also play an important role, and dominate Ohmic resistivity in the more
tenuous and more strongly magnetized disk upper layers
\citep{SalmeronWardle05,SalmeronWardle08,Wardle07}. The linear dispersion properties
of the MRI in the Hall and AD regimes differ substantially from those in the Ohmic regime.
In particular, the inclusion of the Hall effect makes the properties of the MRI depend on the
orientation of the magnetic field \citep{Wardle99,BalbusTerquem01,WardleSalmeron11}.
In the presence of both vertical and azimuthal field, MRI can grow at appreciable rate even
in the limit of infinitely strong AD \citep{KunzBalbus04,Desch04}.

The effects of Ohmic, Hall and AD on the non-linear evolution of the MRI have been
studied separately by various numerical simulations. One of the key questions addressed
by these simulations is that under which conditions can the MRI be sustained. It has been
found that for Ohmic resistivity, MRI can be sustained when the
vertical Elsasser number $\Lambda_z\equiv v_{Az}^2/\eta_O\Omega$ is greater than unity
\citep{Turner_etal07,IlgnerNelson08}, where $v_{Az}$ is the vertical component of the
Alfv\'en velocity, $\eta_O$ is the Ohmic resistivity, and $\Omega$ is the angular frequency
of the disk. Numerical simulations including both the Ohmic resistivity and the Hall term
were performed by \citet{SanoStone02a,SanoStone02b}. They found that the Hall effect
does not strongly affect the saturation level of the MRI, and the $\Lambda_z>1$ criterion
for the MRI to be sustained still holds with the inclusion of the Hall term. Although their
exploration of the Hall parameter was not quite complete \citep{WardleSalmeron11}, these
results are not surprising given the fact that the Hall term is not dissipative.

In our recent paper, \citet{BaiStone11a} (hereafter, BS11), we studied the effect of AD on
the non-linear evolution of the MRI in the strong coupling limit, which applies when the ion
inertia is negligible and the recombination time is much shorter than the orbital time.
This assumption will be justified in this paper in the context of PPDs.
The effect of AD is characterized by the Elsasser number based on AD, denoted by $Am$,
which describes the number of times a neutral molecule ``collides" (i.e., exchanges
much of its momentum) with the ions in an orbital time. The key result of BS11 is
summarized in their Figure 16: MRI can be sustained at any value of $Am$, provided that
the magnetic field is sufficiently weak $\beta\geq\beta_{\rm min}(Am)$, where $\beta$ is the
ratio of gas to magnetic pressure (see Equation (25) of BS11), and $\beta_{\rm min}(Am)$
increases with decreasing $Am$. This relation provides another constraint on the
sustainability of the MRI.

In this paper, we aim at studying the location and extent of the active regions in PPDs,
and predicting the MRI-driven accretion rate in the most realistic manner by incorporating
all the currently available numerical simulation results. We do so by solving a
complex set of chemical reaction network established in \citet{BaiGoodman09}
(hereafter BG09). A single population of dust grains is also included in the network.
Magnetic diffusion coefficients are calculated from the equilibrium abundance of
charged species. A unique feature in our treatment is that we have included the full
dependence of magnetic diffusion coefficients (hence the Elsasser number) on
the magnetic field strength, with the field strength constrained by the results from
non-ideal MHD simulations. This allows us to predict the magnetic field strength and
the accretion rate in PPDs using the least amount of assumptions.
One closely related work is by \citet{Wardle07}, who performed similar chemistry
calculations to obtain magnetic diffusivities of all non-ideal MHD effects with a simpler
reaction network, but the extent of the active layer and accretion rate was not addressed
in detail which may be partly due to the unavailability of numerical simulation results.
Another closely related work to ours is by \citet{PerezBeckerChiang11}, who were
motivated by the accretion problem in transitional disks and the role of tiny grains.
They have considered both Ohmic resistivity and AD, although their adopted
criteria were more simplified and did not account for the role of magnetic field strength.

This paper is structured as follows. We begin by reviewing the derivation of various
non-ideal MHD effects in Section \ref{sec:nimhd}. We describe our chemical reaction
network and calculation of magnetic diffusivities in Section \ref{sec:method}, where we
also discuss our adopted criteria for the MRI-active layer. In Section \ref{sec:result} we
present the results of our fiducial model calculation, where a framework for estimating
the accretion rate and the magnetic field strength is provided. Using this framework,
we study the dependence of accretion rate on various ionization and disk model
parameters in Section \ref{sec:variation}. We summarize and conclude in Section
\ref{sec:conclusion}.

\section[]{Overview of Non-ideal MHD Effects}\label{sec:nimhd}

Non-ideal MHD effects derive from the generalized Ohm's law. In the single-fluid
framework for weakly ionized gas, fluid density $\rho$ and velocity ${\mb v}$ specify
the density and velocity of the neutrals. Let charged species $j$ has particle mass
$m_j$, charge $Z_je$, number density $n_j$, and drift velocity relative to the neutrals
${\mb v}_j$. Charge neutrality condition applies for non-relativistic MHD: $\sum_jn_jZ_j=0$
(note that $Z_j$ can be either positive or negative). Let ${\mb E}'$ be the electric field
in the frame co-moving with the neutrals, while for non-relativistic MHD, the magnetic
field ${\mb B}$ is the same in all frames. In this co-moving frame, the equation of motion
for charged species (whose inertia is negligible) is set by the balance between the
Lorentz force and the neutral drag, given by
\begin{equation}
Z_je({\mb E}'+\frac{{\mb v}_j}{c}\times{\mb B})
=\gamma_j\rho m_j{\mb v}_j\ ,\label{eq:balance}
\end{equation}
where $\gamma_j\equiv\langle\sigma v\rangle_j/(m+m_j)$ with $\langle\sigma v\rangle_j$
being the rate coefficient for momentum transfer between charged species $j$ and the
neutrals, and $m$ is the averaged particle mass of the neutrals.

The relative importance between the Lorentz force and the neutral drag is characterized
by the ratio between the gyrofrequency and the momentum exchange rate
\begin{equation}
\beta_j\equiv\frac{Z_jeB}{m_jc}\frac{1}{\gamma_j\rho}\ .\label{eq:betaj}
\end{equation}
Charged species $j$ is strongly coupled with neutrals if $|\beta_j|\ll1$, and is strongly
tied to magnetic fields when $|\beta_j|\gg1$.

Since the current density is given by ${\mb J}=e\sum_jn_jZ_j{\mb v}_j$. The
generalized Ohm's law can be obtained by inverting equation (\ref{eq:balance}) to
express ${\mb v}_j$ as a function of ${\mb E}'$. The result is
\begin{equation}
{\mb J}=\sigma_O{\mb E}'_\parallel + \sigma_H\hat{\mb B}\times
{\mb E}'_\perp+\sigma_P{\mb E}'_\perp\ ,\label{eq:OhmLaw}
\end{equation}
where subscripts $_\parallel$ and $_\perp$ denote vector components parallel and
perpendicular to the magnetic field ${\mb B}$, and $\hat\ $ denotes unit vector. The
Ohmic, Hall and Pedersen conductivities are \citep{Wardle07}
\begin{equation}
\begin{split}
\sigma_O&=\frac{ec}{B}\sum_jn_jZ_j\beta_j\ ,\\
\sigma_H&=\frac{ec}{B}\sum_j\frac{n_jZ_j}{1+\beta_j^2}\ ,\\
\sigma_P&=\frac{ec}{B}\sum_j\frac{n_jZ_j\beta_j}{1+\beta_j^2}
\end{split}\label{eq:sigma_general}
\end{equation}
respectively. Note that the Hall conductivity depends on the sign of $Z_j$, while
$\sigma_O$ and $\sigma_P$ are always positive since $\beta_j$ has the same
sign as $Z_j$.

The Ohm's law (\ref{eq:OhmLaw}) can be inverted to give the electric field using
current densities, which then leads to the induction equation modified by non-ideal
MHD terms
\begin{equation}
\frac{\pa{\mb B}}{\pa t}=\nabla\times({\mb v}\times{\mb B})
-\frac{4\pi}{c}\nabla\times[\eta_O{\mb J}+\eta_H({\mb J}\times{\hat{\mb B}})
+\eta_A{\mb J}_\perp]\ ,\label{eq:induction}
\end{equation}
where
\begin{equation}
\begin{split}
\eta_O&=\frac{c^2}{4\pi\sigma_O}\ ,\\
\eta_H&=\frac{c^2}{4\pi\sigma_\perp}\frac{\sigma_H}{\sigma_\perp}\ ,\\
\eta_A&=\frac{c^2}{4\pi\sigma_\perp}\frac{\sigma_P}{\sigma_\perp}-\eta_O\ ,\\
\end{split}\label{eq:eta_general}
\end{equation}
are the desired Ohmic, Hall and ambipolar diffusivities as in equation (\ref{eq:induction}),
determined by the microphysics of ion-neutral and electron-neutral collisions, and
$\sigma_\perp\equiv\sqrt{\sigma_H^2+\sigma_P^2}$. Note that only $\eta_O$ is
independent of $B$. The absolute value of these diffusion coefficients determines the
relative importance of the Ohmic, Hall and AD terms.

The most commonly used magnetic diffusivities are obtained by assuming that electrons
and positively charged ions are the only charge carriers, with all ions having the same Hall
parameter. In this case, one can express the conductivities in terms of $\beta_e$ and
$\beta_i$ for electrons and ions respectively, and since $|\beta_e|\gg\beta_i$, one finds
\citep{SalmeronWardle03}
\begin{equation}
\frac{\pa{\mb B}}{\pa t}=\nabla\times({\mb v}\times{\mb B})
-\nabla\times\bigg[\frac{4\pi\eta_e}{c}{\mb J}+\frac{{\mb J}\times
{\mb B}}{en_e}-\frac{({\mb J}\times{\mb B})\times{\mb B}}
{c\gamma_i\rho\rho_i}\bigg]\ .
\label{eq:induction1}
\end{equation}
Physically, because the electrons are the most mobile species in the gas, the magnetic
field is effectively carried by the electrons. Correspondingly, the Hall and AD terms originate
from electron-ion drift and ion-neutral drift respectively. In dense regions with weak magnetic
field, both electrons and ions are well coupled to the neutrals ($1\gg|\beta_e|\gg\beta_i$),
and Ohmic resistivity dominates. In tenuous regions with strong magnetic field, both
electrons and ions are tied to the magnetic field ($|\beta_e|\gg\beta_i\gg1$), and AD
dominates due to the ion-neutral drift. The Hall dominated regime due to the electron-ion
drift lies in between, where electrons are tied to the magnetic field while the ions are
coupled to the neutrals ($|\beta_e|\gg1\gg\beta_i$).  The above formula no longer holds
when a substantial fraction of charged particles are grains \citep{Bai11b}, although it is
widely used for its simplicity and physical clarity. In this paper, we shall adopt the general
expression for magnetic diffusivities (\ref{eq:eta_general}).

The energy dissipation rate associated with the non-ideal MHD effects is given by
\begin{equation}
\dot{E}=\frac{1}{c}{\mc E}_n\cdot{\mb J}=\frac{4\pi}{c^2}(\eta_OJ^2+\eta_AJ_\perp^2)\ ,
\end{equation}
where ${\mc E}_n\equiv(4\pi/c)[\eta_O{\mb J}+\eta_H({\mb J}\times{\hat{\mb B}})
+\eta_A{\mb J}_\perp]$ is the electromotive force associated with the non-ideal MHD
terms. We see that the Ohmic resistivity dissipates the total current (leading to magnetic
reconnection), while AD damps the perpendicular component of the current (via
ion-neutral drag). On the other hand, the Hall effect is not dissipative. It describes the
magnetic diffusion due to drift motion between charged carriers without breaking
magnetic field lines, and is also present in fully ionized plasma.

\section[]{Calculations of Non-ideal MHD Effects in PPDs}\label{sec:method}

As discussed in Section \ref{sec:nimhd}, full assessment of the non-ideal MHD effects
requires knowledge of the number densities of all charged species in PPDs. In this
section, we describe our chemistry calculations to infer magnetic diffusivities in PPDs
and provide a quantitative criterion for judging whether the MRI can operate or not under
the given diffusivities based on results from numerical simulations. Most of the chemistry
calculation procedures are adopted from BG09.

\subsection[]{Disk Model}\label{ssec:disk}

Fiducially, we take the minimum-mass solar nebula (MMSN) model
\citep{Weidenschilling77b,Hayashi81} as our disk model, which is simply constructed
by smearing the required mass for forming the solar system planets into a
smooth distribution. It represents the minimum amount of disk mass required for forming
the solar system planets. The surface mass density and disk temperature are given by
\begin{equation}
\begin{split}
 \Sigma_g&=1700r_{\rm AU}^{-3/2}{\rm\ g\ cm}^{-2}\ ,\\
 T&=280r_{\rm AU}^{-1/2}{\rm \ K}\ ,\\
\end{split}\label{eq:nebula}
\end{equation}
where $r_{\rm AU}$ is the disk radius measured in AU, and the disk is treated as
vertically isothermal. By default, we assume the mass of the protostar to be
$M_*=1M_{\odot}$, with the Keplerian frequency $\Omega=\sqrt{GM_*/r^3}$. The
mean molecular weight of the neutrals is taken to be
$\mu_n=2.34m_H$, from which the sound speed $c_s=\sqrt{kT/\mu_nm_p}$ and disk
scale height $H=c_s/\Omega$ can be found.

In addition, we also consider the solar nebula model proposed by \citet{Desch07}, which
takes into account the recent advances in the planet formation theory (in particular, the
``Nice" model, \citealp{Tsiganis_etal05}), and is much more massive than the MMSN.
The surface density and temperature profiles are given by
\begin{equation}
\begin{split}
 \Sigma_g&=5\times10^4r_{\rm AU}^{-2.17}{\rm\ g\ cm}^{-2}\ ,\\
 T&=150r_{\rm AU}^{-0.43}{\rm \ K}\ ,\\
\end{split}\label{eq:desch}
\end{equation}
where the temperature profile is estimated from \citet{ChiangGoldreich97}.

Submillimeter interferometric observations of the $\sim1$ Myr old Ophiuchus star
forming regions by \citet{Andrews_etal09,Andrews_etal10} have revealed the density
and temperature profiles in the outer regions ($\gtrsim10$ AU, due to limited spatial
resolution) for a sample of PPDs. The surface density for the majority of the disks
appears to match the MMSN value well at $10-20$ AU. Although the fitted density
profile in the outer disk is shallower than the MMSN and the Desch's model (with the
median slope of $-0.9$ rather than $-1.5$ or $-2.2$), the surface density profile of the
inner disk is not well constrained by observations, and both MMSN and the Desch's
model may be viable choices. Furthermore, global calculations of PPD evolution
do indicate much higher surface mass densities in the inner disk than direct
continuation of the observed density profiles to small radii \citep{Zhu_etal10b}.

\subsection[]{Ionization Sources}\label{ssec:ion}

The PPDs are generally too cold for thermal ionization to take place except in the
innermost regions ($<1$ AU, \citealp{Fromang_etal02}). We are interested in regions
with $r\gtrsim1$ AU and consider the following three non-thermal ionization sources.

First, the X-ray ionization from the protostar. Most T-Tauri stars produce strong X-ray
emission due to corona activities (see review by \citealp{Feigelson_etal07}). The X-ray
fluxes are generally variable, with large X-ray flares recurring on time scale of a few
weeks \citep{Stelzer_etal07}. The X-ray emission during the flares is harder than that in
the quiescent state. The time averaged X-ray luminosity is roughly proportional
to stellar mass, and is about $10^{29}$ to $10^{31}$ erg s$^{-1}$ for solar mass stars
\citep{Preibisch_etal05,Gudel_etal07}, with typical X-ray temperature ranging from
$1$-$8$ keV \citep{Wolk_etal05}. We adopt the X-ray temperature $T_X=5$ keV, and
X-ray luminosity $L_X=10^{30}$ erg s$^{-1}$ as our standard model parameters. We
take the ionization rate ($\xi_X^{\rm eff}$) calculated by \citet{IG99}, which takes into
account both absorption and scattering of X-ray photons.
In practice, we use the fitting formula given by BG09 (see their Equation (21))
\begin{equation}
\begin{split}
\xi_X^{\rm eff}=\frac{L_{X,30}}{r_{\rm AU}^{2.2}}&\bigg[4.0\times10^{-11}
e^{-(N_H/N_1)^{0.5}}\\
+&2.0\times10^{-14} e^{-(N_H/N_2)^{0.7}}+ {\rm bot.}\bigg]{\rm s}^{-1}\ ,
\end{split}
\end{equation}
where $L_{X,30}=L_X/10^{30}$erg s$^{-1}$, $N_1=3.0\times10^{21}$ cm$^{-2}$,
$N_2=1.0\times10^{24}$ cm$^{-2}$, $N_H$ denotes the column number density of hydrogen
nucleus from the point of interest to one side of the disk surface, while the ``${\rm bot.}$"
symbol represents the dual terms with $N_H$ being the hydrogen column number density
to the other side of the disk surface. The column number densities $N_1$ and $N_2$ roughly
correspond to column mass density of $1.2\times10^{-2}$ g cm$^{-2}$ and $3.9$ g cm$^{-2}$
respectively.

Secondly, the cosmic-ray (CR) ionization with ionization rate \citep{UN81}
\footnote{See \citet{UN09} for a more refined formula.}
\begin{equation}
\xi_{\rm CR}^{\rm eff}=1.0\times10^{-17}\exp{(-\Sigma/96{\rm g\ cm}^{-2})}
\ {\rm s}^{-1}+{\rm bot.}
\end{equation}
The CR flux is highly uncertain because on the one hand, the flux may be much
higher if a supernova explosion occurs in the vicinity of the protostar, and observations
of the CR flux toward the diffuse cloud $\zeta$ Persei indicate enhanced ionization rate
of $10^{-16}$ s$^{-1}$ \citep{McCall_etal03}; but on the one hand, the CR flux may be
substantially shielded by the stellar wind.

Third, the radioactive decay, primarily the decay of short lived $^{26}$Al, produces
ionization rate of $3.7\times10^{-19}$ s$^{-1}$ with half-life $0.717$ Myr
\citep{TurnerDrake09}. Here we adopt the ionization rate of $10^{-19}$ s$^{-1}$ as
appropriate for disk ages of around $3$ Myr. The radioactive decay is generally too
weak to provide sufficient ionization, but it prevents the midplane of the inner disk from
being completely neutral.

Other possible ionization sources such as energetic protons from disk and stellar corona
\citep{TurnerDrake09} are ignored for simplicity. Although the resulting ionization rate
may exceed X-ray and cosmic-ray ionization by a factor of as large as 40
\citep{PerezBeckerChiang11}, their fluxes are highly variable and uncertain. In our calculations
we also consider $L_X=10^{32}$ erg s$^{-1}$ whose ionization rate is likely to overwhelm this
effect by at least an order of magnitude.

Very recently, \citet{PerezBeckerChiang11b} pointed out another potentially important
ionization source: the far ultraviolet (FUV) radiation from the protostar. FUV photons are
unattenuated by the hydrogen column, and efficiently ionize tracer species of heavy
elements such as C and S with penetration depth of up to $0.1$g cm$^{-2}$. FUV
ionization is not included in our calculation,
while its relative importance
will be discussed in Sections \ref{ssec:rate} and \ref{ssec:var_ion}.

\subsection[]{Grain Size and Abundance}\label{ssec:grain}

The abundance and size distribution of grains in PPDs are crucial for disk chemistry.
Observationally, they are constrained by modeling the spectra energy distribution
(SED) \citep{ChiangGoldreich97,DAlessio_etal98,DullemondDominik04}. Although the
parameters are very degenerate, there have been evidence of grain growth to above
micron size \citep{DAlessio_etal01}, as well as dust settling (to the midplane, which
exhibits as grain depletion, \citealp{Chiang_etal01,DAlessio_etal06,Furlan_etal06,
Watson_etal09}). Results
from mid-infrared spectroscopy from both T-Tauri stars and Herbig Ae/Be stars
\citep{vanBoekel_etal03,vanBoekel_etal05,Przygodda_etal03} also revealed the
presence of micron-sized grains in a substantial fraction of PPDs. In our
chemistry calculation, we adopt a simplified prescription of single-sized, well-mixed
grains, and consider grain sizes of $0.1\mu$m and $1\mu$m. The internal density of
the grains is taken to be $3$ g cm$^{-3}$. Although a size
distribution of grains and some level of grain settling would be more realistic,
our treatment is sufficient to demonstrate the basic physics. Moreover, BG09
considered two populations of grains and found that the two populations behave
essentially independently. They further found that for a range of grain sizes considered
($a=0.01\mu$m to 1$\mu$m), the controlling parameter lies somewhere between the
total grain surface area, and the grain abundance weighted by linear size. For a
population grains with {\it mass} distribution $f(a)$ (with $\int f(a)da=Z=0.01$), one
may conveniently consider
\begin{equation}\label{eq:Sgr}
S\equiv\int_{a_{\rm min}}^{a_{\rm max}}da\bigg(\frac{a}{1\mu{\rm m}}\bigg)^{-3/2}
\frac{f(a)}{0.01}
\end{equation}
as a measure of the chemical significance of dust grains. Our choice of $a=0.1\mu$m
and $a=1\mu$m correspond to $S=32$ and $S=1$ respectively. For a continuous size
distribution of grains, the situation can be more complicated, but one may calculate the
resulting $S$ and use our results as a guide.

Another potentially important ingredient of dust grains is the polycyclic aromatic
hydrocarbon (PAH), which represents the smallest end of grain size distribution. 
More than $50\%$ of the Herbig Ae/Be disks have been detected to have PAH
emission \citep{AckeAncker04}, while the fraction in T-Tauri stars is much smaller
\citep{Geers_etal06,Oliveira_etal10}. The importance of PAHs on the disk conductivity
has been raised by \citet{PerezBeckerChiang11} recently. They argue that PAHs in
T-Tauri disks may be equally abundant as in Herbig Ae/Be disks, but they are
undetected because T-Tauri stars are much less luminous in ultraviolet (UV)
radiation to excite fluorescence emission from the PAHs. The abundance of PAHs in
PPDs is not well determined, but the small size of PAHs means that they can have
very large abundance (which rapidly recombine electrons) without contributing much
to the total dust mass, posing a potential threat to the MRI. In this paper, we do not
include PAHs in our chemical network by assuming they are insufficiently abundant.
However, in our companion paper \citep{Bai11b}, we point out that due to their own
conductivity, charged PAHs can even facilitate the MRI by suppressing the Hall effect
and ambipolar diffusion under certain situations.

\subsection[]{Chemical Reaction Network}\label{ssec:network}

We adopt the complex chemical reaction network described in full detail in
\citet{IlgnerNelson06} and BG09 (see their Section 3). It contains nine elements, 174
gas-phase species and 2083 gas-phase chemical reactions \footnote{We correct the
number of gas-phase reactions in BG09 where it was stated to be 2113.}.
The rate coefficients are taken from the UMIST database \citep{Woodall_etal07}
(except for the ionization reactions, whose rate is given in Table 1 of BG09). We further
include a single population of grains as described in the previous subsection. Grain
related reactions include collisional charging with electrons and ions, adsorption and
desorption of neutrals, and grain collisions. The maximum grain charge is taken to be
$\pm10$ and $\pm30$ elementary charges for $0.1\mu$m and $1\mu$m grains
respectively. This leads to substantial increase of total number of species (to 266 for
$a=0.1\mu$m) and total number of reactions (to 4513 for $a=0.1\mu$m and more than
9000 for $a=1\mu$m) compared with BG. We have made a correction to
\citet{IlgnerNelson06} and BG09 in the calculation of the electron sticking coefficient,
which is described in Appendix \ref{app:sticking} and discussed in Section
\ref{ssec:chemistry}. The calculation of all other reaction rate coefficients remain
unchanged from BG09.

We initialize the chemical reaction network from single-element species, whose
abundance is provided in Table 6 of \citet{IlgnerNelson06}. For the two metal elements
Mg and Fe, we take their abundance relative to hydrogen nucleus to be
$1.0\times10^{-8}$ and $2.5\times10^{-9}$ respectively (same as most calculations in
BG09). The grains are assumed to be uniformly mixed in the disk with $1\%$ in mass.
The large set of stiff ordinary differential equations are evolved by the
fourth-order implicit Kaps-Rentrop integrator with adaptive time stepping described in
\citet{Press_etal92}. Conservation of charge and elemental abundance is enforced at
each step of evolution. The chemical network is evolved for $10^6$ years when
quasi-equilibrium has reached.

\subsection[]{Calculation of Magnetic Diffusivities}\label{ssec:diffusivity}

To formally obtain the conductivities one needs to know the momentum transfer rate
coefficients $<\sigma v>$. For ion-neutral collisions, the neutral atom is induced with
an electrostatic dipole moment as it approaches the ions, the resulting rate coefficient
is approximately independent of temperature, and is inversely proportional to the
reduced mass (\citet{Draine11}, see Table 2.1 and Equation (2.34)):
\begin{equation}\label{eq:ionneutral}
<\sigma v>_i=2.0\times10^{-9}\bigg(\frac{m_H}{\mu}\bigg)^{1/2}
{\rm cm}^3\ {\rm s}^{-1}\ ,
\end{equation}
where $\mu=m_i\mu_n/(m_i+\mu_n)$ is the reduced mass in a typical ion-neutral
collision.

For electron-neutral collisions, we adopt the approximate fitting formula from
\citet{Draine_etal83}, which applies at $T\gtrsim100$K. The dependence of the
rate coefficient on $T$ becomes shallower at lower temperatures due to the
polarization effects, and we adopt
\begin{equation}\label{eq:elneutral}
<\sigma v>_e=8.3\times10^{-9}\times\max\bigg[1\ ,\
\bigg(\frac{T}{100K}\bigg)^{1/2}\bigg]\ {\rm cm}^3\ {\rm s}^{-1}
\end{equation}
as an approximation.

For collisions between neutrals and charged grains, the rate coefficient follows
equation (\ref{eq:ionneutral}) for sufficiently small grains, while
the collision cross section becomes geometric for large grains. Therefore, we have
\begin{equation}
\begin{split}
<\sigma v>_{gr}=\max&\bigg[1.3\times10^{-9}|Z|\ ,\\
1.6\times10^{-7}&
\bigg(\frac{a}{1\mu{\rm m}}\bigg)^2\bigg(\frac{T}{100K}\bigg)^{1/2}\bigg]
\ {\rm cm}^3\ {\rm s}^{-1}\ .
\end{split}
\end{equation}

With these collision rate coefficients, it is then straightforward to evaluate the magnetic
diffusion coefficients ($\eta_O$, $\eta_H$ and $\eta_A$) from equations (\ref{eq:betaj}),
(\ref{eq:sigma_general}) and (\ref{eq:eta_general}).

\subsection[]{Recombination Time}\label{ssec:rcb}

The recombination time $t_{\rm rcb}$ is an important quantity for studying the gas
dynamics in PPDs. If $t_{\rm rcb}$ is much shorter than the dynamical time
($\Omega^{-1}$), as is required in the so called ``strong coupling" limit \citep{Shu91},
local ionization equilibrium would be a good approximation and the magnetic diffusion
coefficients can be directly evaluated from the ionization rate and local thermodynamic
quantities such as density and temperature. This will simplify numerical calculations
considerably because a single-fluid approach is sufficient. In the opposite limit, if
$t_{\rm rcb}$ is much longer than the dynamical time, a more appropriate approach
would be the multi-fluid method.

The recombination time $t_{\rm rcb}$ is not a well-defined quantity when there are
multiple species of ions and when grains are present. Here we propose an effective
recombination time $t_{\rm rcb}^{\rm eff}$ by noticing the fact that resistivity scales
linearly with the abundance of ionized species (especially free electrons, see Section
\ref{sec:nimhd}). After evolving the reaction network to chemical equilibrium, we turn
off the ionization sources and let the system relax (recombine) for a short period of
time ($\ll$1 year). We measure the rate of change of the Ohmic resistivity, and the
effective recombination time is defined as
\begin{equation}
t_{\rm rcb}^{\rm eff}\equiv\frac{\eta_O}{|d\eta_O/dt|}\ .\label{eq:rcb}
\end{equation}
This definition captures the contribution from all recombination channels and is
sensitive to the most rapid recombination processes. It naturally generalizes the
recombination time defined for single ion-electron recombination
$t_{\rm rcb}^{\rm sig}=n_e/(dn_e/dt)\propto1/n_e$.\footnote{One can also define the
effective recombination time based on Hall and ambipolar diffusivities, which gives
similar numbers but may have weak dependence on the magnetic field strength.}

\subsection[]{Criteria for Sustaining the MRI}\label{ssec:criteria}

In weakly ionized disks, non-ideal MHD terms dominate the inductive term if the
Elsasser number is less than $1$, where the Elsasser number based on the Ohmic,
Hall and AD terms are defined as
\begin{equation}
\begin{split}
\Lambda&\equiv\frac{v_A^2}{\eta_O\Omega}\ ,\\
\chi&\equiv\frac{v_A^2}{\eta_H\Omega}\approx\frac{\omega_h}{\Omega}\ ,\\
Am&\equiv\frac{v_A^2}{\eta_A\Omega}\approx\frac{\gamma_i\rho_i}{\Omega}\ .
\end{split}
\end{equation}
Here $v_A=\sqrt{B^2/4\pi\rho}$ is the Alfv\'en velocity, and the approximate equality
in the definition of $\chi$ and $Am$ holds in the grain-free case. The Hall frequency
$\omega_h\equiv eBn_e/\rho c$ is the cutoff frequency of the left polarized Alfv\'en
waves, which can be rewritten to $\omega_h=(n_e/n)(m_i/\mu_n)\omega_{ci}$ with
$\omega_{ci}$ being the ion cyclotron frequency, $n$ being the number density of the
neutrals, $m_i$ being the ion mass.

Non-ideal MHD effects change the linear properties of the MRI substantially when any
of the above Elsasser numbers falls below 1 \citep{BlaesBalbus94,Jin96,Wardle99,
BalbusTerquem01,KunzBalbus04}.
Below we summarize numerical studies on the non-linear evolution of the MRI in these
non-ideal MHD regimes and provide our criteria on the strength and sustainability of
the MRI turbulence, which are crucial to this work.

For the Ohmic resistivity, vertically stratified shearing box simulations have identified that
the border between the MRI active and MRI inactive regions is well described by
$\Lambda=1$ \citep{IlgnerNelson08}, or $\Lambda_z=1$ \citep{Turner_etal07}, where the
vertical Elsasser number $\Lambda_z=v_{Az}^2/\eta_O\Omega$, with $v_{Az}$ being the
vertical component of the Alfv\'en velocity. The former criterion gives slightly thicker active
layer since $\Lambda$ is generally larger than $\Lambda_z$ by a factor of $3-30$. The
difference between these two criteria can be accommodated by
noticing the the fact that transition from the MRI active to MRI inactive regions (i.e., the
``undead zone") is relatively smooth, with the extent of the transition region to be about
$0.5H$ \citep{TurnerSano08}. Therefore, in our calculations, we will simply adopt
$\Lambda\gtrsim1$ as the criterion for the active layer in the Ohmic dominated regime.
It provides an optimistic estimate of the lower boundary of the active layer.

For the Hall effect on the MRI, the only existing non-linear study is performed by
\citet{SanoStone02a,SanoStone02b}. They are motivated by whether the suppression
of the MRI by Ohmic resistivity is affected by the Hall effect and performed simulations
including both the Ohmic and Hall terms. They found that the saturation level of the MRI
is not affected by the Hall effect by much. In particular, the condition for sustained MRI
turbulence is still controlled by the Ohmic Elsasser number and appears to be independent
of the Hall effect. However, the range of the Hall Elsasser number $\chi$ (or $X=2/\chi$ in
their notation) studied by Sano \& Stone is relatively narrow. Whether their conclusion can
be generalized to the Hall dominated regime is still an open question. In this paper, we
tentatively adopt Sano \& Stone's conclusion and ignore the Hall effect
on the sustainability of the MRI, but we also discuss the applicability and potential
caveat about this simplification in Section \ref{ssec:diff}.

The effect of AD on the non-linear evolution of the MRI has been studied by
\citet{HawleyStone98} using a two-fluid approach, applicable when the ionization fraction
is large and the recombination time is long, and by BS11 when the ion inertia is negligible
and when $t_{\rm rcb}$ is much shorter than the orbital period (i.e., the strong coupling
limit). The behaviors of the MRI in the two limits differ significantly. As we will show in
Section \ref{ssec:diff}, the strong coupling limit applies almost everywhere in typical PPDs.
In this regime, BS11 showed that the MRI can be self-sustained for any value of $Am$ as
long as the magnetic field is sufficiently weak. At a given $Am$, the maximum magnetic
field strength is given by
\begin{equation}
\beta_{\rm min}=\bigg[\bigg(\frac{50}{Am^{1.2}}\bigg)^2
+\bigg(\frac{8}{Am^{0.3}}+1\bigg)^2\bigg]^{1/2}\ ,\label{eq:betamin}
\end{equation}
where the plasma $\beta=P_{\rm gas}/P_{\rm mag}$ is the ratio of gas to magnetic
pressure, with $P_{\rm mag}=B^2/8\pi$. Here $\beta$ is not to be confused by the
Hall parameter for charged particles $\beta_j$, which is used only in Section
\ref{sec:nimhd}.

In sum, our adopted criteria for sustained MRI turbulence is
\begin{equation}
\Lambda\geq1\ ,\quad{\rm and}\quad\beta\geq\beta_{\rm min}(Am)\ .\label{eq:criteria}
\end{equation}
Besides the potential uncertainty from ignoring the Hall effect, another uncertainty in
our adopted criteria is that they are mostly based on unstratified simulations with
vertical box size fixed at $H$ (e.g., simulations by Sano \& Stone and BS11). Although
in the Ohmic regime unstratified and stratified simulations tend to yield similar criterion
\citep{Sano_etal98,Fleming_etal00,Turner_etal07}, it is yet to be explored whether the
same situation holds for the case of ambipolar diffusion. In addition, above the MRI
active layer, magnetic dissipation may generate a hot disk corona, which on the one
hand, possesses some stress (although much smaller than that in the active layer,
\citealp{MillerStone00}), and on the other hand, increases $\beta$ in the upper disk.
Despite all these complications, our approach serves as the first step towards more
realistic criteria, and moreover, it is sufficiently simple for illustrating our new method
for estimating the location of the active layer and the accretion rate in Section
\ref{ssec:rate}.

\subsection[]{Accretion Rate and Required Field Strength}\label{ssec:strength}

In the active layer, the MRI generates turbulent stress $T_{r\phi}$ which transports
angular momentum outward, and its strength is usually characterized by the $\alpha$
parameter \citep{ShakuraSunyaev73}, defined as $T_{r\phi}=\alpha\rho c_s^2$.
When MRI is self-sustained, there is a tight correlation between $\alpha$ and the time
and volume averaged magnetic energy, which is again characterized by the plasma
$\beta$ (\citealp{HGB95}, BS11)
\begin{equation}
\alpha\approx\frac{1}{2\beta}\ .\label{eq:alphbeta}
\end{equation}
Note that this relation holds only in the MRI-active layer.

If accretion in PPDs is solely driven by the MRI in the active layer, a simple relation can be
derived connecting the accretion rate $\dot{M}$ to the magnetic field strength (BG09). For
steady state accretion, conservation of angular momentum demands
\begin{equation}\label{eq:csv}
\dot{M}\Omega r^2=2\pi r^2\int_{-\infty}^{\infty}dzT_{r\phi}
\approx2\pi r^2\int_{\rm active}dzT_{r\phi}\ ,
\end{equation}
where the last integral is performed across the active layer. Although the dead zone has also
been shown to transport angular momentum by non-axisymmetric density waves launched
from the active layer \citep{FlemingStone03,OishiMacLow09}, which is a generic process in
shear turbulence \citep{HeinemannPap09a,HeinemannPap09b}, its contribution is only a
small fraction of that from the active layer, and can be safely ignored.
From equation (\ref{eq:alphbeta}), we have
$T_{r\phi}=\alpha P_{\rm gas}\approx P_{\rm mag}/2$ in the active layer,
and the above equation turns to
\begin{equation}
\dot{M}\approx\int_{\rm active}dz\frac{B^2}{8\Omega}\ .\label{eq:acrate}
\end{equation}
This is a very useful formula for accretion rate estimation and is quite general for MRI
driven angular momentum transport in both ideal MHD or non-ideal MHD regimes.
It will be used extensively in Sections \ref{ssec:rate} and Section \ref{sec:variation}.

In turn, for MRI driven accretion, one can estimate the strength of the magnetic field
given the accretion rate. Let the thickness of the active layer be denoted by $h_a$
for each side of the disk, the integral over the vertical height can be replaced by a
factor of $2h_a$, which leads to
\begin{equation}
\langle B^2\rangle\approx4\dot{M}\Omega/h_a\ ,\label{eq:B2constrain}
\end{equation}
where the bracket $\langle\cdot\rangle$ means vertical averaging. To obtain a more
quantitative estimate of the field strength applicable to PPDs, we consider the MMSN
around a $1M_{\odot}$ protostar. The thickness of the active layer should generally
be on the order of the disk scale height $h_a\approx H=c_s/\Omega$, and we obtain
\begin{equation}
\langle B\rangle\approx2\sqrt{\dot{M}\Omega/H}\approx1.0\dot{M}_{-8}^{1/2}
r_{\rm AU}^{-11/8}\ {\rm G}\ ,\label{eq:Bconstrain}
\end{equation}
where $\dot{M}_{-8}=\dot{M}/10^{-8}M_{\odot}$ yr$^{-1}$. We note that in obtaining
the above relation, only the temperature profile (to estimate the disk scale height) of the
MMSN model is used (which is more reliable than the surface density profile). This
relation states that for MRI driven angular momentum transport, strong magnetic field is
needed for fast accretion.

For typical accretion rate of $10^{-8}M_{\odot}$ yr$^{-1}$, the implied magnetic field
strength is strong, and is in fact close to the equipartition strength for the
active layer. Assuming the column density of the active layer $\Sigma_a$ to be $10$ g
cm$^{-1}$ (as comparable to the penetration depth of the X-ray ionization), the
equipartition field strength is
\begin{equation}
B_{\rm equi}\approx\sqrt{8\pi}\bigg(\frac{\Sigma_a}{h_a}c_s^2\bigg)^{1/2}
\approx2.3\Sigma_1^{1/2}r_{\rm AU}^{-7/8}\ {\rm G}\ ,
\end{equation}
where $\Sigma_1=\Sigma_a/10$ g cm$^{-2}$. The simple analysis here shows that to
the limiting factor for MRI to drive accretion in PPDs is not only the level of ionization,
but also the strength of the magnetic field. A quantitative manifestation of this effect
will given in Section \ref{ssec:rate}.

\section[]{Active Layer in the Fiducial Model}\label{sec:result}

We refer our fiducial model to the MMSN disk with X-ray luminosity of the protostar
being $10^{30}$erg s$^{-1}$ and with cosmic-ray ionizations. Variations to the fiducial
model are discussed in the next section. Other parameters are fixed at values specified
in the previous section in all calculations.

Using the fiducial model, we run four calculations at two disk radii $1$AU and $10$AU,
with and without grains ($0.1\mu$m). In each calculation, we evolve the chemical
network at fixed radius and scan from the disk midplane up to 5 disk scale heights.
At each point, we extract the number density of all species from the end of the evolution
($10^6$ years) and calculate the magnetic diffusivities as a function of the magnetic field
strength. Various aspects of the results are discussed in the following subsections.

\subsection[]{Chemistry}\label{ssec:chemistry}

\begin{figure}
    \centering
    \includegraphics[width=90mm]{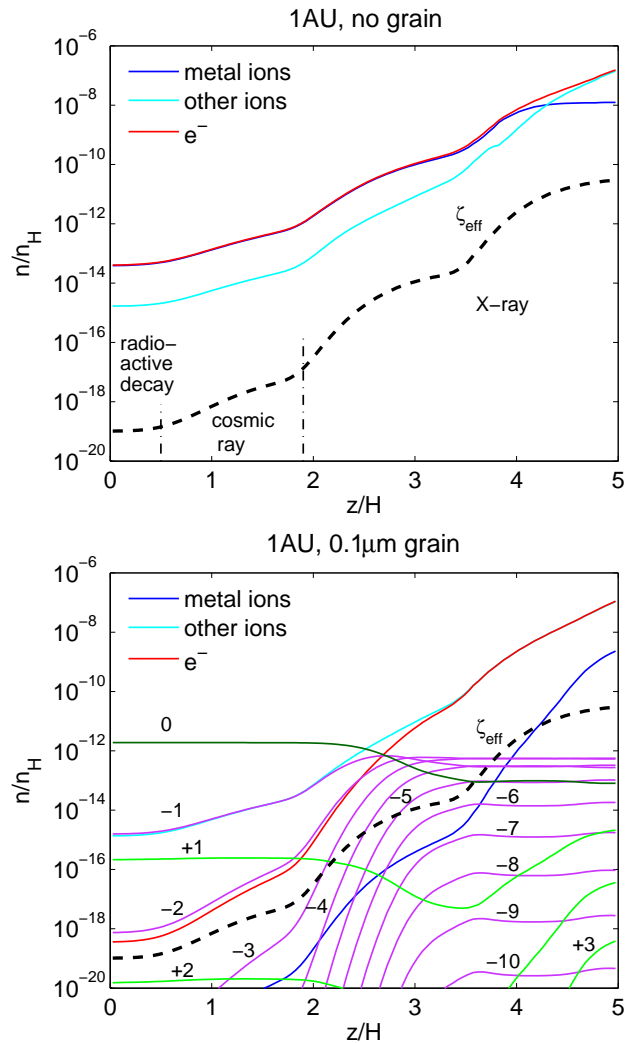}
  \caption{Fractional abundance of electrons (red), metal ions (blue) and other ions
  (cyan) relative to hydrogen nuclei as a function of height ($z$) above the midplane
  in our fiducial model at 1AU. Also shown is the ionization rate (bold dashed) as a
  function of $z$, divided into three segments where the dominant sources of ionization
  are labeled. Upper panel: calculation without grains. Lower panel: calculation with
  well mixed $0.1\mu$m grains with $1\%$ in mass. Abundance of positively charged
  (green), neutral (dark green) and negatively charged (magenta) grains are labeled by
  their elemental charge $Z$.}\label{fig:chemistry}
\end{figure}

In Figure \ref{fig:chemistry}, we show the vertical density profile of various chemical
species normalized to the number density of the hydrogen nuclei for calculations at $1$ AU.
The most important quantity is the ionization fraction $x_e\equiv n_e/n_H$, as plotted in red,
which largely determines the strength of the non-ideal MHD effects (Section \ref{sec:nimhd}).
In addition, we plot the profile of the ionization rate. This figure is to be compared with
Figures 3 and 6 in \citet{Wardle07}. It is clear that the ionization fraction is extremely small
($\ll10^{-6}$ in general), hence the inertia of the charged particles is negligible compared
with the inertia of the neutrals, justifying the first requirement of the strong coupling limit.

The main driving force of chemical evolution is the ionization reactions. In the Figure,
there are several ``steps" in the vertical profile of the ionization rate that are associated
with transitions to different ionization regimes, as described in Section \ref{ssec:ion}.
These ``steps" also make the vertical profile of the ionization fraction $x_e$ exhibit similar
features. At the uppermost layer, the ionization rate is the largest and is dominated by
direct X-ray ionization from the protostar, with a very small column density of about $0.01$
g cm$^{-2}$. Slightly deeper down, the ionization is still dominated by the X-rays, but is
mainly due to the Compton scattered X-ray photons from the upper layer, with penetration
depth of about $4$ g cm$^{-1}$. Further deeper, cosmic-ray ionization with penetration
depth of about $100$ g cm$^{-1}$ takes over, while around the midplane, radioactive
decay dominates. Comparing the result to Figure \ref{fig:diffusivity} shown in the next
subsection reveals that the ionization rate provided by radioactive decay is generally too
small to produce sufficient ionization in PPDs, and can be safely ignored for the purpose
of estimating the extent of the active layer and dead zone.

In the grain-free calculation, we see that the ionization fraction primarily overlaps with
the metal abundance at $z\lesssim4H$, indicating that the metals are the dominant
electron donor, which has been shown in many previous works (e.g.,
\citealp{Fromang_etal02,IlgnerNelson06}, BG09). Above $4H$, essentially all the metal
atoms are ionized, and the main electron donor is taken over by other ions. The ionization
fraction from our calculation differs from that in \citet{Wardle07} mainly because we use a
more complex (and presumably more realistic) chemical network.

The inclusion of well-mixed dust grains with $a=0.1\mu$m dramatically reduces the
ionization fraction. At the midplane, $x_e$ is reduced by 5 orders of magnitude as
compared with the grain-free case. The reduction factor is still significant but smaller in
the upper layers up to $z\gtrsim5H$.
With grains, the role for metals as the main electron donor is suppressed because the
recombination of metal ions is facilitated by grains, consistent with \citet{Wardle07}.

The reduction of electron density by grains has another consequence: when the electron
abundance falls substantially below the grain abundance, the ions and grains take over from
the electrons to play a decisive role on the conductivity. In the chemistry calculation shown
in the bottom panel of Figure \ref{fig:chemistry}, this corresponds to regions close to the
midplane with $|z|\lesssim2.5H$. The grain-free formula (\ref{eq:induction1}) for magnetic
diffusivities no longer holds in this regime: the resistivity $\eta_O$ is smaller than $\eta_e$
due to contribution from ions and grains. In addition, $Am$ is no longer independent of
magnetic field strength (as can be traced from Figure \ref{fig:constrain_fid}), and becomes
larger in stronger field. This effect and its significance is discussed in full detail in
\citet{Bai11b}.

One difference between our calculation and the calculations by \citet{Wardle07} is that we
have considered the electron sticking probability $s_e$. This probability $s_e$ was simply
taken to be $1$ in their calculation. In \citet{PerezBeckerChiang11}, $s_e$ was fixed at
$0.1$ for PAHs, and $1$ for normal grains, while in \citet{Okuzumi09}, $s_e$ was fixed at
0.3 for all grain sizes.
However, because electron is much lighter than the grain surface atoms, energy transfer by
inelastic collisions with grains is inefficient and the electron may have a large chance to
escape. The derivation of electron sticking coefficient with {\it neutrals} grains has been
performed by \citet{Nishi_etal91}, who showed that $s_e$ decreases with increasing
temperature, from about $1$ at zero temperature to about 1-2 orders of magnitude less
than unity at a few hundred K. \citet{IlgnerNelson06} and BG09 adopted this formula in their
chemistry calculations, but they erroneously used the same formula for both neutral and
charged grains. A generalized derivation of the electron sticking coefficient to include grain
charges is given in Appendix \ref{app:sticking}, where we find that the sticking
coefficient for more positively charged grains is progressively smaller than that for
negatively charged grains. This is mainly because the electron is accelerated more as it
approaches the more positively charged grains, making it more difficult to get rid of the
excess energy for adsorption.

The inclusion of the sticking coefficient in grain-electron collisions reduces the
electron recombination rate with grains, leading to less dramatic reduction of
electron abundance and conductivity. Our test runs indicate that that the ionization
fraction calculated with and without including the electron sticking probability can
differ by up to a factor of $10$ in certain regions. The electron sticking probability also
affects the grain charge distribution. For example, in Figure \ref{fig:chemistry}, the
mean grain charge in the disk upper layers from our calculation is about $-2.5$
elementary charge rather than around $-8$ in \citet{Wardle07}'s calculation. Our new
result on the charge dependence of $s_e$ implies that grains tend to be (slightly)
more positively charged, although this is a relatively weak effect and is more relevant
for sub-micron grains.

\subsection[]{Magnetic Diffusivities and Recombination Time}\label{ssec:diff}

\begin{figure*}
    \centering
    \includegraphics[width=160mm,height=140mm]{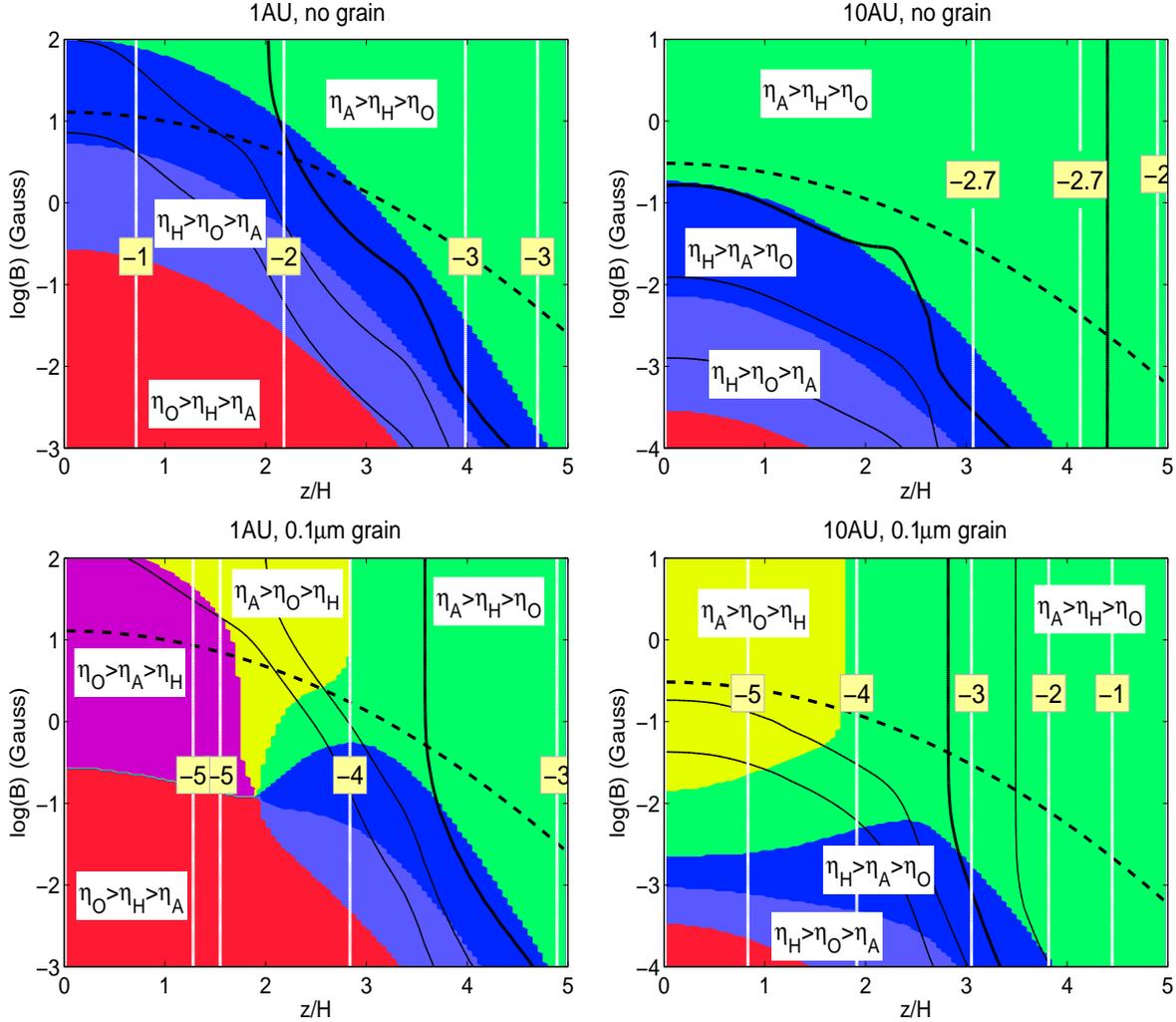}
  \caption{Regimes of non-ideal MHD effects in fiducial PPD models plotted as
  contours in the plane of disk height and magnetic field strength. The left (right)
  two panels correspond to 1 (10) AU of a MMSN model, and the upper (lower) two
  panels correspond to the case without grains (with $1\%$ of well mixed
  $0.1\mu$m grain). Different regimes of non-ideal MHD effects are painted with
  different background colors. {\it Red and magenta}: Ohmic resistivity dominated;
  {\it Dark and light blue}: Hall effect dominated; {\it Green and yellow}: AD
  dominated. Subdivisions of the color scheme are indicated in the plots.
  {\it Black contours} show constants of the Elsasser number $\Lambda_{\rm tot}$
  which is increased by factors of 10 from bottom left to upper right, with the
  $\Lambda_{\rm tot}=1$ contour marked in bold. {\it The bold dashed black line}
  indicates where the magnetic pressure equals to the gas pressure ($\beta=1$).
  White {\it vertical lines} correspond to contours of constant effective recombination
  time $t_{\rm rcb}^{\rm eff}$, labeled by $\log_{10}(\Omega t_{\rm rcb}^{\rm eff})$.}
  \label{fig:diffusivity}
\end{figure*}

The abundance of all charged species from the chemistry calculation is used to
evaluate the magnetic diffusivities using equations (\ref{eq:sigma_general}) and
(\ref{eq:eta_general}), and the results are illustrated in Figure \ref{fig:diffusivity}.
Similar to the figures shown in \citet{Wardle07}, we mark different magnetic diffusion
regimes with different filling color. Six magnetic diffusion regimes are considered
depending on the relative order among $\eta_O$, $\eta_H$ and $\eta_A$. It is clear
that Hall and AD becomes more and more important at smaller density (surface layer
and large disk radii) and larger magnetic field strength, as expected ($\eta_H$ scales
as $B/n_e$ while $\eta_A$ scales as $B^2/n_e$). The addition of grains dramatically
changes the pattern in the figure at $z\lesssim3H$, this is because grains carry
most of the negative charge instead of electrons (due to the extremely low ionization
rate) and the magnetic diffusivity is largely determined by the less mobile ions and
grains. In this region, different choices of chemical reaction networks can make a big
difference in the resulting magnetic diffusivity pattern. At disk upper layers
($z\gtrsim3H$), grains play a less important role on the pattern of magnetic
diffusivities painted in the Figure since free electrons overwhelms the grains, although
the ionization fraction is still affected by the grains.

White vertical lines in Figure \ref{fig:diffusivity} show the contours of constant
effective recombination time (equation (\ref{eq:rcb})). We see that in all of the four
cases, the recombination time is at least one order of magnitude smaller than the
dynamical time scale\footnote{Note that our definition of $t_{\rm rcb}^{\rm eff}$ in
equation (\ref{eq:rcb}) captures the most rapid recombination process. It is
typically shorter than the chemical equilibrium time estimated in
\citet{PerezBeckerChiang11}, which is sensitive to the slowest chemical processes}.
This result looks counterintuitive since the recombination time may be expected to
be smaller in the disk upper layers due to the low gas density. However, this is
compensated by the enhanced electron abundance near the disk surface (as
$t_{rc}\sim1/nx_e$). Together with the extremely low ionization level in PPDs, this
result demonstrates that the strong coupling limit applies in essentially most regions
of typical PPDs, and it justifies that single fluid treatment of the gas dynamics in
PPDs is generally sufficient. In particular, the single fluid simulations by BS11 on
the effect of AD on the MRI is directly relevant to PPDs, while two-fluid simulations
by \citet{HawleyStone98} are not quite applicable.

Moreover, we emphasize that our conclusion that $\Omega t_{\rm rcb}^{\rm eff}\ll1$
is obtained by using the complex chemical network. The usage of a simple
network such as the \citet{OD74} model can lead to different conclusions: At $1$ AU
without grains, we find that $t_{\rm rcb}$ is about one order of magnitude longer when
calculated with the simple network, and becomes longer than the dynamical time at
$|z|\lesssim H$. This is because of the lack of recombination channels in the simple
network, and is relevant to the ``revival" of the dead zone by turbulent mixing of free
electrons from the active layer to the midlane, as seen in multi-fluid simulations with
a co-evolving simple chemical reaction network
\citep{Turner_etal07,TurnerSano08,IlgnerNelson08}. However, with a more realistic
chemical network, the reactivation of the dead zone by turbulent mixing would appear
less likely to occur, because the turbulent eddy time is comparable to the dynamical
time \citep{FromangPap06,Turner_etal06,Carballido_etal11} and most free
electrons would be swallowed by the more rapid recombination process before being
mixed down to the midplane. Therefore, the density profile of all charged species in
PPDs should be close to local ionization equilibrium, which justifies our adopted
criteria in Section \ref{ssec:criteria}.

In Figure \ref{fig:diffusivity}, we also plot contours of constant Elsasser number. Here
we define the Elsasser number based on $\eta_{\rm tot}$ as
\begin{equation}
\Lambda_{\rm tot}\equiv\frac{v_A^2}{\Omega\eta_{\rm tot}}\ ,
\end{equation}
where $\eta_{\rm tot}=\sqrt{\eta_O^2+\eta_H^2+\eta_A^2}$, and it measures the
importance of all non-ideal MHD effects as a whole. The Elsasser number is
in general larger in upper right and smaller in lower left, with the
$\Lambda_{\rm tot}=1$ contours plotted in bold black lines. Non-ideal MHD terms
dominate the induction term when $\Lambda_{\rm tot}<1$, with the properties of the
MRI significantly modified. In addition, we plot the contour of $\beta=1$ in bold dashed
line. Before applying our criteria (\ref{eq:criteria}), which will be the subject of the next
subsection, we note that the bold solid and bold dashed lines may serve as
{\it rough} boundaries enclosing the MRI active region. We see that in all cases the
dominant non-ideal MHD processes in these regions are the Hall effect and AD.

As we discussed in Section \ref{ssec:criteria}, we have ignored the Hall effect on
the sustainability of the MRI turbulence. Here we discuss the potential caveats from
this simplification. We are most interested in the boundary between MRI active and
inactive regions. The lower boundary set by $\Lambda=1$ is relatively well constrained
based on the study of \citet{SanoStone02b}, although more numerical study in the Hall
dominated regime is still necessary since it is generally the case that $\eta_H>\eta_O$
near the lower boundary. The main uncertainty comes from the upper boundary
($\beta\geq\beta_{\rm min}(Am)$), which is based on simulations of BS11 that include
only AD. From Figure \ref{fig:diffusivity}, we see that AD is indeed the dominant
non-ideal MHD effect close to the line of $\beta=1$ at 10 AU. Therefore, our criterion
$\beta\geq\beta_{\rm min}(Am)$ should provide a relatively reliable upper boundary
on the strength of the magnetic field for MRI to operate. At 1 AU with grains, AD is
also likely to be the main limiting factor on the magnetic field strength, while in
the grain-free case, the location of the upper boundary is likely to be in the Hall
dominated regime, and our criterion may require modification.

\subsection[]{Active Layer and Accretion Rate}\label{ssec:rate}

\begin{figure*}
    \centering
    \includegraphics[width=160mm,height=140mm]{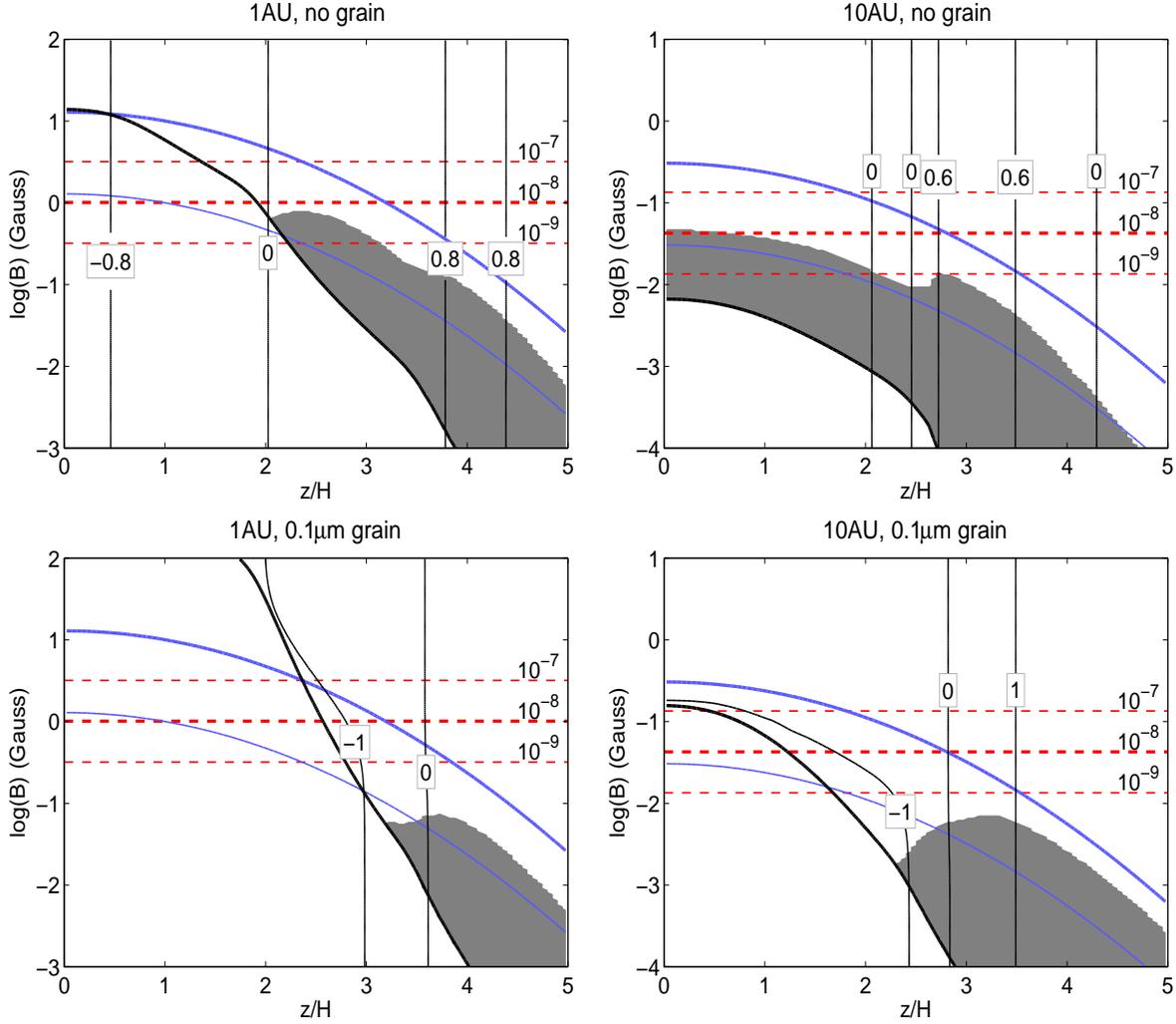}
  \caption{Similar to Figure \ref{fig:diffusivity}, but for constraints on the MRI permitted
  region in PPDs. The bold solid contours correspond to the Ohmic Elsasser number
  $\Lambda=1$, while the thin solid curves show the contours of constant $Am$ labeled
  by $\log_{10}(Am)$, terminated at $Am=0.1$. Blue bold and thin lines mark the plasma
  $\beta=1$ and $\beta=100$ respectively. Permitted regions for the MRI in
  PPDs based on criteria (\ref{eq:criteria}) are painted in gray. The red dashed lines
  indicate the required field strength in the MRI permitted region corresponding to
  accretion rate of $10^{-7}$, $10^{-8}$ and $10^{-9}M_{\odot}$ yr$^{-1}$, respectively,
  based on equation (\ref{eq:Bconstrain}).}\label{fig:constrain_fid}
\end{figure*}

Using the profile of magnetic diffusivities in the previous subsection, we estimate the
location of the active layer by applying our criteria (\ref{eq:criteria}). In Figure
\ref{fig:constrain_fid}, we show the $\Lambda=1$ contour (black bold solid) as well as
contours of constant $Am$ (black thin solid) in the $z$-$B$ plane similar to Figure
\ref{fig:diffusivity}. The former (Ohmic resistivity) controls the lower boundary of the
active layer, while its exact location is determined by the strength of the magnetic field:
lower for strong field and higher for weak field. This is because the Ohmic Elsasser
number $v_A^2/\eta_O\Omega$ is proportional to $B^2$, and the disk is better ionized
(smaller $\eta_O$) in the upper layer. The upper limit on the magnetic field strength
characterized by $\beta$ (blue solid) is controlled by AD, where smaller $Am$ value
requires weaker magnetic field. The value of $Am$ above the lower boundary is
generally greater than $0.1$, and reaches up to $10$ in the disk upper layers, which
agrees with PAH-free model calculations in \citet{PerezBeckerChiang11}, and the
minimum value of $\beta$ ranges from about $10$ to $800$. The two constraints
combined together give the MRI permitted region in the $z$-$B$ plane at a given
radius the PPD, as painted in gray.

We see that in all of the four cases, maintaining an MRI active layer is possible if the
magnetic field is not too strong. That is, the required field strength to overcome the
Ohmic resistivity in the disk upper layer is generally much weaker than the
equipartition field. This is partly due to the fact that $\eta_O\propto1/x_e$ decreases
toward the disk surface more rapidly than density (at least for $z<4H$), making the
$\Lambda=1$ contour steeper than the contours of $\beta=$ constant. Also, $Am$ in
the disk upper layer is generally greater than $1$, where the required field reduction
from equipartition strength is only moderate. The MRI permitted region in the $z$-$B$
plane shifts to disk upper layers when sub-micron grains are included because of
increased $\eta_O$. It extends toward the disk midplane as one moves to disk outer
regions because the reduction of disk surface density slows the recombination
process and allows ionization
photons/particles to penetrate deeper. At $10$ AU without grains, even the disk
midplane becomes active for field strength of a few times $0.01$G. These results
are all consistent with previous chemistry calculations.

For MRI-driven accretion, the required magnetic field strength at various accretion
rates from equation (\ref{eq:Bconstrain}) is also shown in Figure \ref{fig:constrain_fid}
(red dashed). We see that in the grain-free calculations, the maximum field strength in
the MRI permitted region corresponds to $\dot{M}\approx10^{-8}M_{\odot}$ yr$^{-1}$,
while in the presence of sub-micron grains, the maximum accretion rate is dramatically
reduced to well below $10^{-9}M_{\odot}$ yr$^{-1}$.
The reduction is mostly due to the retreat of the lower boundary: the $\Lambda=1$
contour shifts upward, reducing the maximum allowed field strength because of reduced
gas pressure; in addition, $Am$ becomes smaller, reducing the allowed field strength
further.

We note that most previous calculations either adopt the magnetic Reynolds number
Re$_M\equiv c_s^2/\eta_O\Omega=100$ as the boundary between the active layer
and the dead zone, which implicitly assumes $\beta=100$ in the entire disk
(\citealp{Fromang_etal02,IlgnerNelson06}; BG09;
\citealp{PerezBeckerChiang11,PerezBeckerChiang11b}), or adopt the Elsasser number
criterion, but assume some constant magnetic field strength \citep{TurnerDrake09}.
From Figure \ref{fig:constrain_fid}, the Re$_M=100$ criterion corresponds to using the
intersection between the bold solid line ($\Lambda=1$) and thin blue line ($\beta=100$)
as the boundary. We see that this criterion roughly applies in the grain-free case, but may
substantially overestimate the active column when grains are present (i.e., the critical
Re$_M$ should be larger than 100). In other words, the reduction of the active layer
column density by grains is more serious than previously estimated, complementing
our discussion in the previous paragraph.

Our graphical illustration of the maximum accretion rate based on the approximate
formula (\ref{eq:Bconstrain}) can be improved with a more quantitative calculation to
give an absolute upper limit of the accretion rate. This upper limit can be obtained
from equation (\ref{eq:acrate}), where the integral is performed across the entire disk
height that the MRI permitted region extends to, with $B$ chosen to be the
maximum value determined by $\beta_{\rm min}(Am)$. For the MMSN disk model, we
have
\begin{equation}
\begin{split}\label{eq:acrate_max}
\dot{M}_{\rm max}&\approx\frac{c_s}{8\Omega^2}\int_{\rm active}B_{\rm max}^2
\frac{dz}{H}\\
&\approx0.98\times10^{-8}r_{\rm AU}^{11/4}\int^{z>0}_{\rm active}
\bigg(\frac{B_{\rm max}}{\rm G}\bigg)^2\frac{dz}{H}\ M_{\odot}\ {\rm yr}^{-1}\ .
\end{split}
\end{equation}
For the four calculations in Figure \ref{fig:constrain_fid}, we find the maximum accretion
rate (in unit of $M_{\odot}$ yr$^{-1}$) to be $4.5\times10^{-9}$, $1.5\times10^{-8}$
for the grain free model at 1 AU and 10 AU respectively, $3.8\times10^{-11}$ and
$2.5\times10^{-10}$ for well-mixed sub-micron grain model at 1 AU and 10 AU. As
expected, they agree with the estimate from our graphical illustration.

Our equation (\ref{eq:acrate}) provides an convenient way for estimating the accretion
rate from the magnetic field strength, but it does not predict as a priori the field strength in
the disk. Here we consider the following thought experiment and hypothesis from which
we propose a way to predict the strength of the magnetic field in the active layer hence
the accretion rate.

Suppose the initial magnetic field strength in the disk is sufficiently weak (say,
$\beta=10^5$ at the disk midplane). According to Figure \ref{fig:constrain_fid}, the MRI
first develops in the very upper layer of the disk. The MRI amplifies the magnetic field, 
and consequently, both the upper and lower boundaries of the active layer moves
toward the midplane because of increased field strength\footnote{This is somewhat
related to the recurrent growth and decay of the MRI turbulence as a result of field
amplification and resistive damping observed by \citet{Simon_etal11}, who adopted a
constant resistivity profile. In real situations with a rapidly increasing resistivity toward
the disk midplane, the transition may be a more smoothed process.}. The MRI ultimately
saturates into turbulence, and turbulent stirring maintains roughly constant field strength
across the active layer (e.g., \citealp{OishiMacLow09}). We further hypothesize that the
magnetic field in global disks is able to self-organize into a configuration to maximize
the rate of angular momentum transport by the MRI turbulence. While unjustified due
to the lack of systematic global studies (most global simulations of the MRI such as
\citet{FromangNelson06} and local stratified simulations have zero net vertical flux,
which leads to relatively weak turbulence), this hypothesis predicts the magnetic field
strength and accretion rate that lie on the larger side than in reality (i.e., an optimistic
estimate).

Based on the discussion above, we predict the accretion rate as follows. We scan over
the strength of the magnetic field $B$ in the active layer ($B$ is assumed to be
constant across the active layer), where for each $B$, we calculate $\dot{M}$ using
equation (\ref{eq:acrate}). The optimistically predicted accretion rate is the maximum $\dot{M}$
from the scan. For the four calculations in Figure \ref{fig:constrain_fid}, the predicted accretion
rates (in unit of $M_{\odot}$ yr$^{-1}$) are $2.7\times10^{-9}$, $7.7\times10^{-9}$ for grain free
models at 1 AU and 10 AU respectively, and $2.3\times10^{-11}$ and $1.3\times10^{-10}$ for
well-mixed sub-micron grain models at 1 AU and 10 AU. They are about half the value of the
upper limits listed before.

While the framework we just described is fully general, the results reported in this paper are
subject to a few caveats due to simplifications in the adopted disk and ionization models that
may result in small uncertainties up to a factor of a few. These uncertainties are discussed
below and can be overcome by adopting more realistic disk and ionization models that takes
into account radiative transfer, thermodynamics and magnetic pressure support to yield more
reliable estimate of the accretion rate.

We have assumed constant vertical temperature profile in our estimate of the accretion rate,
while in reality, the upper layer of the disk is heated by both the X-ray photons which are
responsible for the ionization \citep{Glassgold_etal04,GortiHollenbach04} and the MRI itself
(BG09, \citealp{HiroseTurner11}), raising the temperature above the midplane temperature by
a factor of a few. This effect increases the gas pressure, hence plasma $\beta$, and allows
stronger magnetic field in the active layer, which may lead to higher $\dot{M}$ than our predicted
value by a factor of a few.

We have assumed that the gas density profile is Gaussian in our calculations. In reality,
magnetic pressure plays more important role in the hydrostatic equilibrium at the disk
surface, and local isothermal stratified shearing-box simulations by \citet{MillerStone00}
show deviation from Gaussian density profile with gas density at the disk surface a factor of
up to 10 higher than our adopted Gaussian model. This may promote accretion by allowing
stronger magnetic field near the surface. However, since the ionization rate depends on the
column gas density from the disk surface, this effect is compensated by the reduced
ionization rate and enhanced recombination rate (at given disk height), giving only order
unity corrections to the predicted accretion rate. In fact, the predicted accretion rate does not
sensitively depend on the density distribution across disk height, and the approximate
calculations by \citet{PerezBeckerChiang11b} that are free from the vertical density
distribution are generally consistent with ours within order unity.

Simulations by \citet{MillerStone00} also indicates non-zero stress in the magnetic corona, but
the $\alpha$ value is generally one order of magnitude smaller than that in the disk. The coronal
contribution to the accretion rate is about $\alpha P_{\rm c}h_c$, with the coronal gas pressure
$P_{\rm c}$ orders of magnitude smaller than that in the disk midplane / active layer, and the
coronal thickness $h_c$ on the order of a few disk scale height (not much thicker than that of the
active layer). Therefore, for our purpose, accretion in the magnetic dominated disk corona can
be safely neglected.

Recently, \citet{PerezBeckerChiang11b} called for attention to the role of the FUV
ionization, which produces much higher ionization fraction than other ionization sources
(making $Am\gtrsim1000$) with penetration depth of about $0.01$ g
cm$^{-2}\lesssim\Sigma_{FUV}\lesssim0.1$ g cm$^{-2}$. We note that in the MMSN
model, $0.1$ g cm$^{-2}$ corresponds to $z\approx3.8H$ at 1 AU and $z\approx2.9H$
at 10 AU, which are very upper layers in the disks. Comparing with Figure
\ref{fig:constrain_fid} we see that at such heights, the equipartition field strengths in the
active layer correspond to $\dot{M}$ of a few times $10^{-10}M_{\odot}$ yr$^{-1}$
and a few times $10^{-9}M_{\odot}$ yr$^{-1}$ respectively, which set the upper limit
for the MRI accretion rate driven by FUV ionization. The upper limit is a factor of a few
below our predicted $\dot{M}$ in the grain-free case, but may well exceed the
predicted rate when small grains are included. 

In sum, we have provided a general framework for estimating the location of the active
layer as well as the accretion rate at a given location of PPDs. The most important
feature is that it explicitly incorporates the dependence on the magnetic field strength,
with the field strength estimated by our physically motivated hypothesis. 
Furthermore, the simulation based relation $\alpha=1/2\beta$ allows us to provide an
optimistic estimate as well as a robust upper limit of the accretion rate, without involving
additional assumptions about the value of $\alpha$. This framework will be extensively
used for parameter study of the accretion rate in the next section, and can be generalized
with more realistic criteria and disk models in the future.

\section[]{Parameter Study}\label{sec:variation}

\begin{figure*}
    \centering
    \includegraphics[width=160mm]{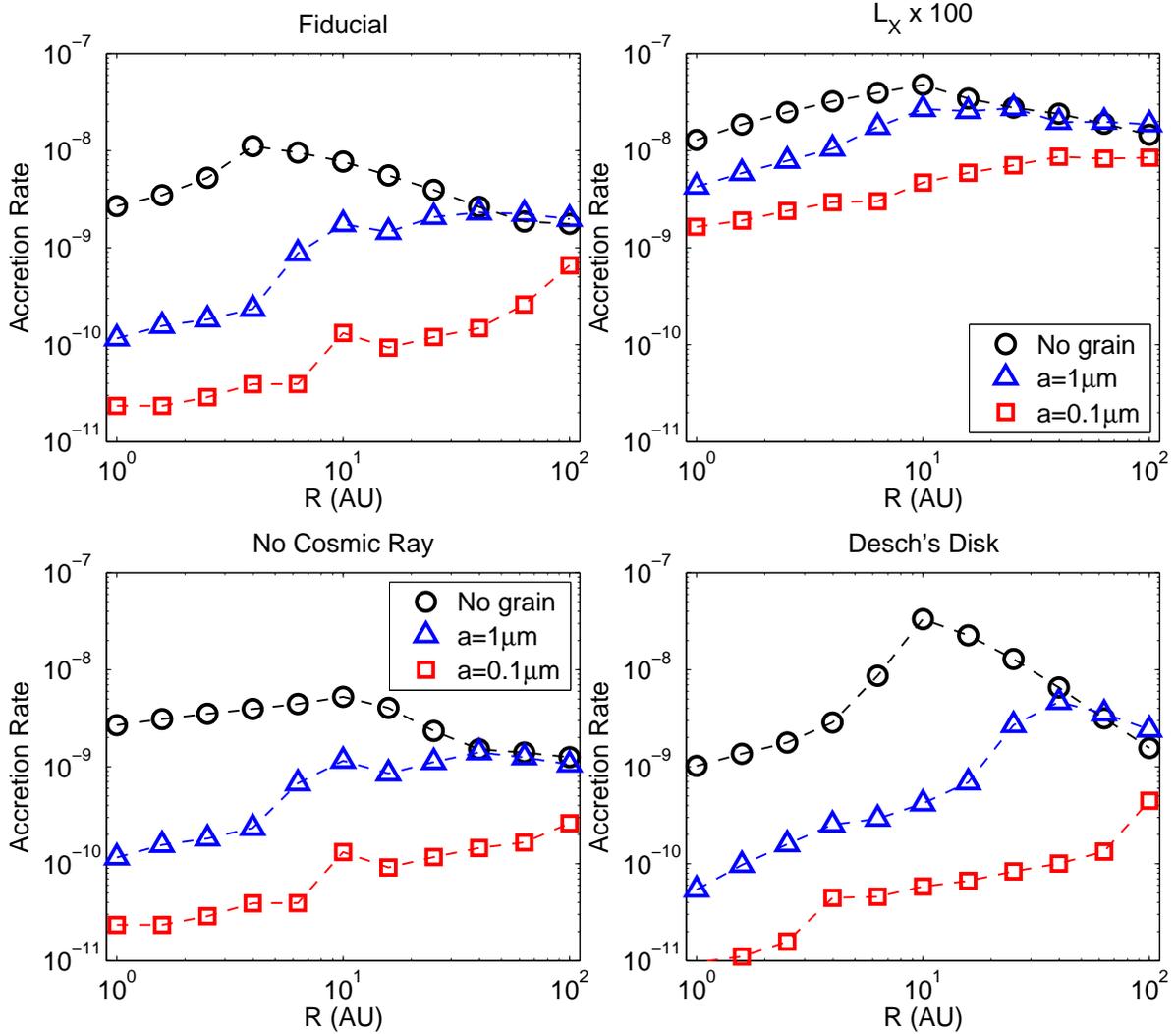}
  \caption{The optimistically predicted accretion rate (in unit of $M_{\odot}$ yr$^{-1}$) as a
  function of disk radius for our fiducial model (upper left): MMSN disk with cosmic ray and X-ray
  ionization, model with X-ray luminosity 100 times higher (upper right), model without
  cosmic-ray ionization (lower left) and model with the Desch's disk (lower right). In each
  panel, we show results for calculations in the grain-free case (black circles), with well-mixed
  $1\mu$m grains (blue triangles) and with $0.1\mu$m grain (red squares). }\label{fig:acrate_all}
\end{figure*}

In this section, we perform a series of chemistry calculations at different disk radii from
$1$AU to $100$AU and explore a number of different model parameters including grain
size, ionization rate and disk mass.
In each series of the parameter study, we derive the MRI-driven accretion rate $\dot{M}$ as
a function of disk radius using the method illustrated in the previous section, in parallel with
the predicted magnetic field strength. The results are shown in Figures \ref{fig:acrate_all}
and \ref{fig:beta_all}, with detailed description and discussion given in the subsections
that follows. Obviously, using a fixed disk model, the predicted $\dot{M}$ is not
necessarily constant as a function of disk radius, which would lead to non-steady accretion
if MRI is the only driving mechanism. On the other hand, the predicted spatially varying
$\dot{M}$ implies that modifications to the adopted disk model is necessary for steady
state accretion. Detailed modeling of the disk structure to match the steady state
accretion is beyond the scope of this paper, but the trend can be readily obtained from
our studies.

\subsection[]{The Effect of Grain Size}\label{ssec:var_gr}

We begin by considering our fiducial model: MMSN disk with both cosmic ray ionization
and X-ray ionization ($L_X=10^{30}$erg s$^{-1}$). In the upper left panel of Figure
\ref{fig:acrate_all} we show the predicted $\dot{M}$ as a function of disk radius. We
have considered the grain-free case as well as models with $1\mu$m and $0.1\mu$m
well-mixed grains with mass fraction of $1\%$. Again, we remind the reader
that our predicted accretion rate is optimistic.

In general, the predicted $\dot{M}$ increases with disk radius $r$ in the inner disk, and
decreases with $r$ in the outer disk. In the former case, the disk midplane
is ``dead" and accretion is layered. Since $\dot{M}\propto\alpha\Sigma_{\rm active} T/\Omega$
(where $\Sigma_{\rm active}$ is the column density of the active layer),  assuming both
$\Sigma_{\rm active}$ and $\alpha$ to be constant, one obtains $\dot{M}\propto r$ for the
MMSN temperature profile. In reality, as we discussed in Sections \ref{ssec:criteria} and
\ref{ssec:rate}, $\alpha$ is controlled by $Am$ (which is about $1$-$10$ in the active layer),
while the thickness of the active layer depends on the mutual effects of Ohmic resistivity
and AD, neither assumptions strictly hold. In Figure \ref{fig:acrate_all}, we see that the
increase of $\dot{M}$ with disk radius in the inner disk is somewhat slower than $r$.

When the entire disk becomes active, Ohmic resistivity becomes essentially irrelevant in
constraining the accretion rate (see the upper right panel of Figures \ref{fig:diffusivity} and
\ref{fig:constrain_fid} at $r=10$ AU for reference). The accretion rate is mainly controlled by
the disk surface density, temperature, and the magnetic field strength
($\dot{M}\propto\alpha\Sigma T/\Omega$) through AD. For MMSN,
$\Sigma T/\Omega\propto r^{-1/2}$, which gives the main source of accretion rate reduction.
The predicted profile of $\dot{M}$ is also affected by the radial profile of the $\alpha$ parameter
determined by $Am$. 

In the grain-free calculation, the entire disk becomes active at $r\gtrsim4$ AU in our fiducial
model, and the predicted $\dot{M}$ is between $10^{-9}$ and
$10^{-8}M_{\odot}$ yr$^{-1}$ at all disk radii considered.  The reduction of the accretion rate
caused by dust grains is substantial. At 1 AU, $\dot{M}$ is reduced by about two orders
of magnitude in the presence of $0.1\mu$m grains, while for $1\mu$ grains the reduction is
more than one order of magnitude. The entire disk becomes active for $r\gtrsim70$ AU and
$r\gtrsim10$ AU respectively in the above two cases. By and large, the predicted $\dot{M}$
increases with disk radius and flattens out at large radii as discussed before. The rise in
predicted $\dot{M}$ in the $0.1\mu$m case toward $100$ AU is mainly because the
cosmic-ray ionization takes over the X-ray ionization (see next subsection).

We find an interesting fact that at very large disk radii ($\gtrsim40$ AU), calculations with
$1\mu$m grains lead to comparable or slightly higher $\dot{M}$ than those without
grains. By carefully checking the equilibrium abundance of all chemical species, we find
that in the grain-free calculations, ionization level near the disk midplane is determined by
metal abundance, while the number density of other ions is much smaller (see Figure
\ref{fig:chemistry} for example). In the presence of grains, metals are depleted at low
temperature (see equation (26) of BG09) due to adsorption onto grain surfaces, and the
dominant ion component becomes species such as H$_3^+$ and HCO$^+$ (and others)
depending on the gas density. The
suppression of metals is one of the reasons for the reduction of ionization level, but it
appears that at very low density and temperature, slightly higher ionization level can be
achieved due to non-metal ions. Nevertheless, we also note that due to uncertainties in
the reaction rate coefficients in the UMIST database, the equilibrium abundance of
various chemical species such as HCO$^+$ can be uncertain up to a factor of $\sim4$
\citep{Vasyunin_etal08}.

We see that even in the absence of dust grains, the optimistically predicted $\dot{M}$ is
only comparable or below $10^{-8}M_{\odot}$ yr$^{-1}$, which is about the median value
of the observed accretion rate in T-Tauri stars \citep{Hartmann_etal98,Sicilia_etal05}.
In the presence of dust grains, the accretion rate that can be driven from the MRI is reduced
dramatically. The situation is particularly serious for T-Tauri disks, since the gas has to enter
through the inner disk to accrete. Given the fact that our choice of parameters are typical,
these results pose strong challenge on the effectiveness of the MRI in driving rapid accretion
in PPDs. Below we further explore other effects to see whether higher MRI-driven accretion
rate can be achieved.

\subsection[]{The Effect of Ionization Rate}\label{ssec:var_ion}

Because of the uncertainty in the amount of cosmic-rays, which may be shielded by the
stellar winds, we also perform a series of chemistry calculations without cosmic-ray
ionization, and the results are shown in the lower left panel of Figure \ref{fig:acrate_all}.
We note that besides deeper penetration depth, the cosmic-ray ionization rate generally
exceeds the scattering component of the X-ray ionization rate beyond about $25$ AU,
and one might expect that cosmic-ray ionization makes significant contributions to our
predicted $\dot{M}$ in the outer disk. However, this does not seem to be the case.

In the grain-free case, we find that the the transition radius larger than which the entire disk
becomes active shifts from about $4$ AU in the fiducial model to about $10$ AU in the
absence of cosmic rays. Nevertheless, the predicted $\dot{M}$ does not change by much
throughout the disk (only slightly reduced). In the presence of grains, the
transition radius shifts outward to about $25$ AU for the $1\mu$m case, and is above
$100$ AU for the $0.1\mu$m case. Again, our predicted $\dot{M}$ is also slightly
reduced (generally within a factor of $2$) but still similar to the fiducial results. To explain,
we note that $\dot{M}$ largely depends on the magnetic field strength in the
active layer. Although X-rays do not penetrate as deep as cosmic rays to activate the disk
midplane in the outer disk, the permitted magnetic field strength in the active layer turns
out to be similar to the case with cosmic rays. According to equation (\ref{eq:acrate}), 
given similar field strength, $\dot{M}$ is then determined by the {\it geometric}
thickness of the active layer, which is on the order of $H$ and is not sensitive to
whether the active layer extends to the disk midplane or not.

To examine the role of X-ray ionization, we further consider a model with $L_X=10^{32}$ erg
s$^{-1}$, 100 times larger than our fiducial X-ray luminosity. Although such X-ray luminosity
is unusually high for T-Tauri stars \citep{Gudel_etal07}, it provides an upper limit on the
accretion rate that can be driven by X-ray ionization. Alternatively, one might also view this
unrealistic choice of high X-ray luminosity as an incorporation of other possible but uncertain
strong ionization sources such as energetic protons from protostellar activities
\citep{TurnerDrake09}. We see from the upper right panel of Figure \ref{fig:acrate_all} that
the predicted $\dot{M}$ is much higher than our fiducial model.
The rise in $\dot{M}$ is mainly due to deeper penetration in the inner disk ($\lesssim5$
AU) and due to reduced AD in the outer disk. Accretion rate in grain-free calculations reaches
a few times $10^{-8}M_{\odot}$ yr$^{-1}$, which is close to the upper limit of the observed
accretion rate. Also, comparing the three curves with their counterparts in the fiducial plot (upper
left) indicates that increasing X-ray luminosity is more efficient in raising accretion rate when
small grains are present. In sum, it appears that for the MRI to explain the observed rate of
accretion in PPDs, much stronger ionization rate than fiducial is needed.

We can also compare our results with the predicted $\dot{M}$ by FUV ionization shown
by \citet{PerezBeckerChiang11b}\footnote{Two order-unity effects may lead
\citet{PerezBeckerChiang11b} to overestimating the accretion rate by a factor of a few
compared with ours: (a) Their equation (6) underestimates the density in the disk upper layer
by a factor of $\gtrsim3$ in a Gaussian density profile, thus promoting higher ionization; (b)
Their calculations correspond to the maximum possible accretion rate (\ref{eq:acrate_max}),
while our predicted rate is about 2 times smaller (see Section \ref{ssec:rate}).}. With the FUV
penetration depth of $\lesssim0.1$ g cm$^{-2}$, they predict $\dot{M}$ of a few times
$10^{-10}M_{\odot}$ yr$^{-1}$ at 1 AU and it increases roughly linearly with radius since
accretion only proceeds in the upper surface layer with roughly constant surface column
density. Comparing with their Figure 5 we see that FUV ionization is likely to be a comparably
important ionization source as X-rays (while they drive accretion in different layers),
and in the presence of small grains, FUV ionization makes significant contributions to the total
MRI-driven accretion rate in the inner disk, and could dominate other ionization sources in the
outer disk.

\subsection[]{The Effect of Disk Mass}\label{ssec:var_disk}

We explore the role of disk mass by repeating our calculations using the Desch's disk
model, while the adopted ionization rates remain fiducial, and the results are shown in
the lower right panel of Figure \ref{fig:acrate_all}. We see that at 1 AU, where the
Desch's disk has about $30$ times more surface density than the MMSN disk, the
predicted $\dot{M}$ for all three cases (with and without grains) are smaller than the
fiducial results. In the grain-free case, the predicted $\dot{M}$ rapidly increases with
radius in the inner disk, and peaks at about $r=10$ AU where the entire disk becomes
active. The peak rate exceeds the MMSN model at the same location by a factor of
several. The predicted rate falls off towards the outer disk more rapidly than that in the
MMSN disk, which is essentially because the surface density of Desch's disk decreases
with radius more rapidly. With the grains, the predicted $\dot{M}$ generally increases with
disk radius until saturation (the entire disk becomes active), and the corresponding
accretion rate in the outer disk is comparable or slightly exceeds that in the MMSN model.
Finally, the rates approach the fiducial results at around $100$ AU simply because the
surface densities in the two disk models are comparable.

Our results suggest that given the same ionization sources, a more massive disk leads
to slower accretion if the accretion is layered, while it is able to sustain faster accretion if
the entire disk is active. The reason is that at the same column density from the disk surface
(hence the same ionization rate), the gas density is higher in a more massive disk (as one
can check with the error function). Therefore, the recombination rate is enhanced, reducing
the ionization fraction hence $\dot{M}$ roughly as $\rho^{-1/2}$. On the other hand, higher
disk mass in the disk permits stronger magnetic pressure in the active layer, and dominates
the former effect when the entire disk column is active.

We have seen that the predicted $\dot{M}$ by the MRI turbulence generally increases
with radius in the inner disk, and decreases with radius in the outer disk, with the transition
radius depending on the dust content. This result is obviously inconsistent with steady state
accretion. Further evolution would lead to pile-up of mass towards the inner disk, with the
density profile in the outer disk becoming more flattened. According to the results in this
subsection, the pile-up of mass in the inner disk would lead to slower rather than faster
accretion, which further enhances the mass pile-up. Therefore, the fact that $\dot{M}$
decreases with disk surface density in the inner disk leads to a runaway pile-up of mass,
which may be unsustainable. This result further support the conclusion that mechanisms
other than the MRI is needed to drive rapid accretion through the inner disk.

\subsection[]{Predicted Magnetic Field Strength}\label{ssec:field}

\begin{figure*}
    \centering
    \includegraphics[width=160mm]{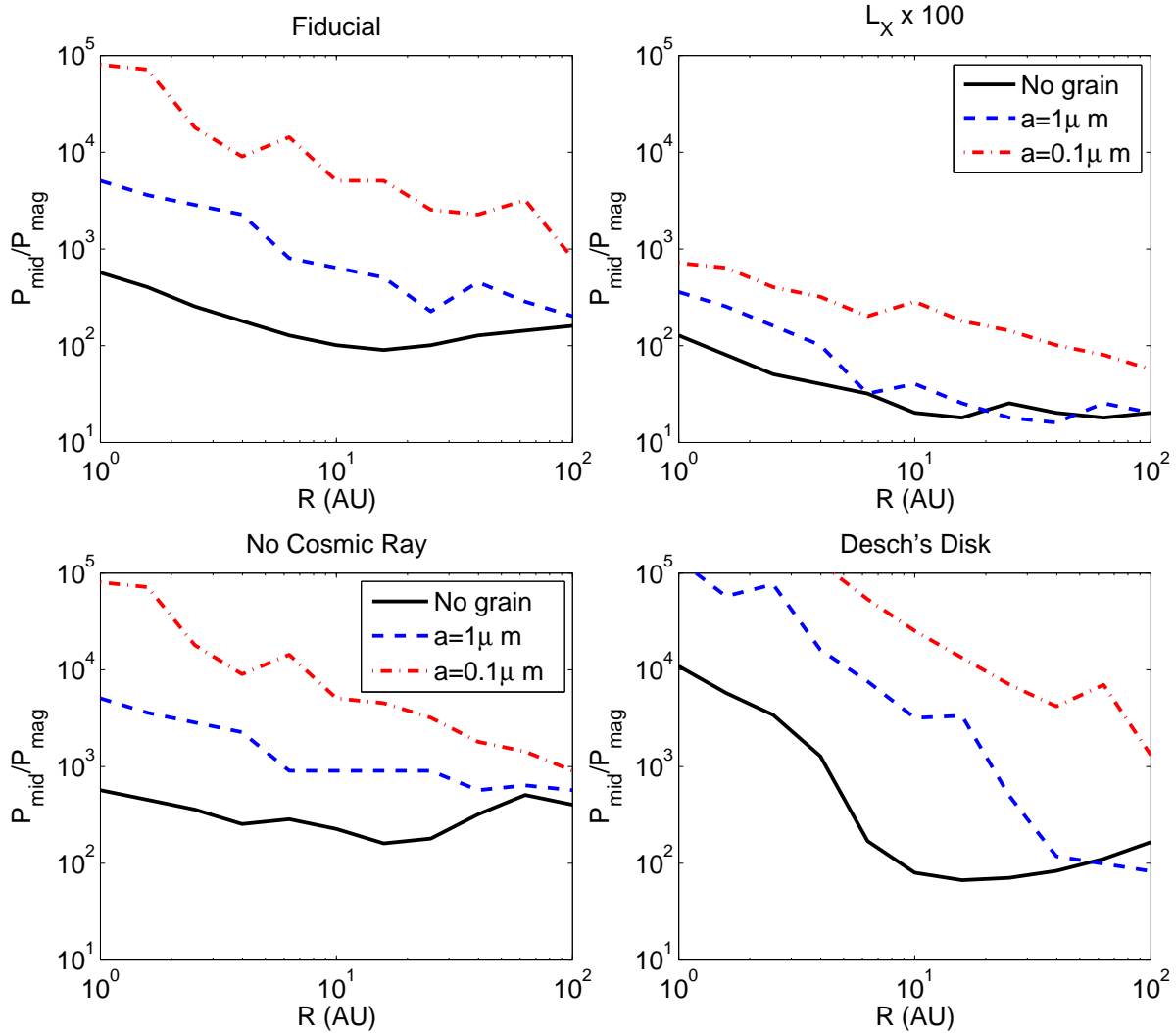}
  \caption{The ratio of midplane gas pressure to the optimal magnetic pressure at the
  active layer as a function of disk radius for our fiducial model
  (upper left): MMSN disk with cosmic ray and X-ray ionization, model with X-ray
  luminosity 100 times higher (upper right), model without cosmic-ray ionization (lower
  left) and model with the Desch's disk (lower right). In each panel, we show results
  for calculations in the grain-free case (black solid), with well-mixed $1\mu$m grains
  (blue dashed) and with $0.1\mu$m grain (red dash-dotted). }\label{fig:beta_all}
\end{figure*}

In Figure \ref{fig:beta_all}, we show the (optimistically) predicted magnetic field strength
in terms of the ratio of midplane gas pressure to the magnetic pressure in the active
layer $\beta_s\equiv P_{\rm mid}/P_{\rm mag}$. This is a physically more convenient
quantity than the absolute magnetic field strength and provides a guide to numerical
modelers. The magnetic field strength can be obtained by
\begin{equation}
\begin{split}
B&=13\beta_s^{-1/2}r_{\rm AU}^{-13/8}\ {\rm G}\qquad ({\rm MMSN})\\
B&=72\beta_s^{-1/2}r_{\rm AU}^{-1.94}\ {\rm G}\qquad ({\rm Desch})\\
\end{split}
\end{equation}
Note the steep dependence of the magnetic field strength on disk radius.

Because the active layer resides in the upper disk, with lower gas pressure hence
smaller magnetic field, $\beta_s$ is generally much larger than $1$. We see that
$\beta_s$ spans a large range between about $100$ and $10^5$, depending on the
surface density, ionization rate and disk radius. The predicted field strength is
generally stronger (smaller $\beta_s$) in the grain-free case since the active layer
resides higher in the disk surface in the presence of grains, with lower gas pressure
hence smaller magnetic field. The field strength can be raised significantly by strong
ionization (upper right panel), accompanied by increased accretion rate as discussed
in Section \ref{ssec:var_ion}. For the massive disk model (lower right), $\beta_s$ in
the inner disk is much higher than the fiducial case because the active layer resides
at a higher position than in the fiducial model due to the finite ionization penetration depth.

It has been predicted that the presence of ordered magnetic field in PPDs can lead to
grain alignment with the magnetic field and produce polarized emission in the millimeter
continuum \citep{ChoLazarian07}. Grain alignment is expected in the outer disk, where
radiative torque dominates thermal collisions. Recently, however, \citet{Hughes_etal09}
reported non-detection of polarized emission using arcsecond-resolution Submillimeter
Array (SMA) polarimetric observations. One suspected reason has to do with the magnetic
field strength, which has to be above some critical field strength for the alignment to occur.
For $10-100\mu$m grains, \citet{Hughes_etal09} calculated the critical field strength to be
$10-100$mG at the location of $50-100$ AU (corresponding to the angular resolution
of the SMA for the observed sources). However, at such distances, our predicted field
strength is less than $2$mG (taking $\beta_s=100$). Therefore, the non-detection of
polarized emission is, in fact, consistent with the presence of the MRI, while if polarized
emission were detected, the implied magnetic field would be too strong for the MRI to
operate in the outer disk.

\section[]{Summary and Discussion}\label{sec:conclusion}

We have explored the non-ideal MHD effects in protoplanetary disks (PPDs) using
chemistry calculations with a complex reaction network including both gas-phase
and grain-phase reactions. Cosmic-ray and X-ray ionization processes are included
with standard prescriptions. The equilibrium abundance of all charged species are
used to calculate the full conductivity tensor, from which diffusion coefficients for
Ohmic resistivity, Hall effect and ambipolar diffusion (AD) are evaluated. One major
finding from our chemistry calculations is that the recombination time is much shorter
than the orbital time in essentially all regions in PPDs, no matter grains are present
or not. Together with the extremely low level of ionization in PPDs, this verifies the
applicability of the ``strong coupling" limit, and the gas dynamics in PPDs can be well
described in a single-fluid framework with magnetic diffusion coefficients in non-ideal
MHD terms given from chemical equilibrium. In particular, turbulent mixing of chemical
species (especially electrons) can be compensated by the rapid recombination process.
Additionally, we have updated the calculation of the sticking coefficient in electron-grain
collisions where grain charge is taken into account. The electron sticking coefficient can
deviate substantially from $1$ which has a strong effect on the electron recombination
rate and grain charge distribution, and should be treated with care.

Using the magnetic diffusion coefficients from the chemistry calculation, we estimate
the location and extent of the regions in PPD models where the MRI can operate
to drive disk accretion (i.e., the active layer). Our adopted criteria are based on the
Ohmic Elsasser number $\Lambda$, where $\Lambda\gtrsim1$ is required for the
active layer as been shown by previous simulations (e,g.,
\citealp{Turner_etal07,IlgnerNelson08}), and the AD Elsasser number $Am$, where
$\beta\gtrsim\beta_{\rm min}(Am)$ is required for sustained turbulence from the most
recent study by BS11. We have ignored the Hall effect based on the study by
\citet{SanoStone02b}, although the Hall regime is still yet to be more carefully
explored with numerical simulations. Unlike most previous studies, we have considered
the dependence of the diffusion coefficients (translated to the Elsasser number) on the
magnetic field strength and show that the magnetic field strength can be a main limiting
factor on the extent of the active layer because of AD. Our study shows that the
conventional magnetic Reynolds number criterion with Re$_M\geq100$ for the MRI
to operate may substantially overestimates the column density of the MRI active layer
in the presence of small grains.

We provide a general framework for estimating the upper limit of the MRI-driven
accretion rate. Furthermore, we hypothesize that the MRI amplifies the magnetic field to
maximize the accretion rate, from which we are able to make optimistic predictions on
the accretion rate as well as the magnetic field strength in PPDs. Using this framework,
we run a series of chemistry calculations with different model parameters to study the
location and extent of the MRI active layers in PPDs, and to study the dependence of
MRI-driven accretion rate $\dot{M}$ on various parameters. The main results are summarized
as follows.

\begin{enumerate}
\item Active layer always exists. Its upper boundary is set by AD, requiring the field
strength not to be too strong. The lower boundary is generally determined by the Ohmic
resistivity and the magnetic field strength, with the latter connected to the field strength
in the upper boundary.

\item The predicted $\dot{M}$ increases with disk radius in the inner disk where
accretion is layered, and flattens or decreases with radius in the outer disk when the
disk midplane becomes active. The transition radius depends on grain abundance, and
is $4-10$ AU in the grain-free case and above $50$ AU in the presence of well-mixed
sub-micron grains.

\item For our standard model, the predicted $\dot{M}$ is a few times
$10^{-9}M_{\odot}$ yr $^{-1}$ in the grain-free case, and is reduced by one to two
orders of magnitude in the presence of sub-micron grains in the inner disk. Reduction
of $\dot{M}$ by grains is less significant in the outer disk.

\item The MRI-driven accretion rate is sensitive to the protostellar X-ray luminosity but
insensitive to the deeper-penetrating cosmic-ray ionization rate. Extremely strong X-ray
luminosity or additional strong ionization source with sufficient grain depletion is needed
to achieve accretion rate of $\sim10^{-7}M_{\odot}$ yr$^{-1}$.

\item In the inner disk where accretion is layered, the predicted $\dot{M}$ increases with
disk radius, but decreases with increasing disk surface density. This situation would lead
to runaway mass pile-up in the inner disk if the MRI were the only mechanism for driving
accretion in the inner region of PPDs.

\item The midplane gas pressure is generally a factor of 100 to $10^5$ times higher than the
predicted magnetic pressure in the active layer. The ratio is smaller for stronger ionization
and higher for larger disk mass or in the presence of small grains. The predicted magnetic
field strength in the outer disk is consistent with the non-detection of polarized emission
resulting from grain alignment.
\end{enumerate}

Our results place strong constraints on the effectiveness of the MRI-driven accretion
in PPDs, especially for the inner disk with radius of about $1-10$ AU where accretion is
likely to be layered. There are two main difficulties concerning points 3 and 5 above:
(a) The optimistically predicted accretion rate under standard model prescriptions is
about one order of magnitude or more smaller than the typical observed value;
(b) MRI-driven accretion would lead to the runaway pile-up of mass in the inner disk.
The first difficulty might be alleviated by incorporating additional (but uncertain)
ionization processes such as energetic protons from the protostellar activities. The
second difficulty does not appear to be easily reconciled, since external ionization
sources always lead to layered accretion. The accumulation of mass may trigger
the gravitational instability in the early phase of PPDs, which may further lead to
outburst and episodic accretion as FU Ori-like events \citep{Zhu_etal09b}. In later
phases, mass accumulation may simply leads to very large surface density in the inner
disk \citep{Zhu_etal10b}. Since the inner disk ($\lesssim10$ AU) is smaller than the
resolution of the currently available sub-millimeter observations \citep{Andrews_etal09},
whether mass accumulation occurs may be distinguishable in future observations
such as ALMA.

One possible resolution to the above difficulties may be achieved by a magnetized wind
from the disk surface \citep{BlandfordPayne82}. The wind could be launched from the
disk surface where the magnetic field is too strong for the MRI to operate, transporting
angular momentum vertically by the Maxwell stress via ordered magnetic fields. Non-ideal
MHD effects has been incorporated into wind models \citep{WardleKoenigl93,Teitler11},
and under favorable conditions, wind-driven accretion rate can become substantial
\citep{Konigl_etal10,Salmeron_etal11}. It has been proposed that disk wind may co-exist
with the MRI \citep{Salmeron_etal07}, while they operate at different vertical locations.
Current wind models for PPDs generally have a number of unconstrained free parameters,
and further investigation (especially by numerical simulations) is needed to assess
the effectiveness of disk wind on PPD accretions.

Our results are also applicable in the gas dynamics of transitional disks, where the MRI is
likely to be responsible for driving accretion from the outer disk \citep{ChiangMurrayClay07},
with the inner gap / hole opened by multiple planets that guide the gas streams through
\citep{PerezBeckerChiang11,Zhu_etal11}. Our predicted accretion rate in the outer disk
with $r\gtrsim10$ AU is on the order of $10^{-9}M_{\odot}$ yr$^{-1}$, with relatively weak
dependence on the dust content compared with the inner disk. The result is consistent with
the range of observed accretion rate \citep{Najita_etal07}, thus MRI alone appears sufficient
to account for the accretion rate in transitional disks. This conclusion can also be achieved
with far UV ionization instead of X-rays \citep{PerezBeckerChiang11b} .

Our study of the MRI active layer in PPDs represents a first attempt to estimate the
MRI-driven accretion rate that is based on the most up-to-date non-ideal MHD
simulations. In the mean time, it is limited by the predictive power of the simulations
themselves. In particular, the non-linear properties of the MRI in the Hall dominated
regime is still unexplored. Moreover, the AD simulations by BS11 are unstratified,
and whether their obtained criterion for sustained MRI turbulence holds remains to
be verified in stratified simulations. In conclusion, our study calls for further progress
in non-ideal MHD simulations, and in turn, more realistic criteria for sustained MRI
turbulence can be easily incorporated into our general framework after future
simulations.

\acknowledgments

X.-N.B. is grateful to Daniel Perez-Becker for valuable comments to the original manuscript
that lead to the corrections of an important code bug and some conclusions of the paper, to
James Stone, Jeremy Goodman, Eugene Chiang and an anonymous referee for carefully
reading the manuscript with helpful comments and suggestions, to Daniel Perez-Becker and
Eugene Chiang for sharing their preprint on the importance of FUV ionization, to Meredith
Hughes, Geoffroy Lesur, Subu Mohanty, Raquel Salmeron, Neal Turner and Zhaohuan Zhu
for useful discussions, and to Roy van Boekel and Greg Herczeg for organizing the Ringberg
workshop for Transport Processes and Accretion in YSOs which led to improvements in this
work.

\appendix

\section[]{Sticking Coefficient in Electron-Grain Collisions and Grain Charging}
\label{app:sticking}

The collision of an electron with a grain particle does not necessarily lead to absorption
because the electron has excess energy as it reaches the grain surface. The absorption
probability $P_e$ depends on the size of the grain $a$, grain charge $Z$, and the
electron energy $E$. The sticking coefficient is defined as the absorption probability
averaged over a thermal distribution of electron energies, which is a function of
temperature $T$
\begin{equation}
s_e(a,Z,T)=\frac{\int_0^\infty \sigma(E)Ee^{-E/kT}P_e(E)dE}
{\int_0^\infty \sigma(E)Ee^{-E/kT}dE}\ ,\label{eq:stickcoeff}
\end{equation}
where $\sigma(E)=\sigma(a,Z,E)$ is the electron-grain cross section is given by
\citet{DraineSutin87} (see their Section II). The task here is to calculate
$P_e(E)=P_e(a,Z,E)$.

The sticking probability in the case of electron collision with neutral grains was derived
in the Appendix B of \citet{Nishi_etal91}. Below we generalize their derivation to include
grain charge.

Let $E_0$ be the initial energy of the electron (at infinity), $D$ be the depth of the
potential well between the electron and the grain due to the electric polarization
interaction (on the order of a few eV). The kinetic energy of the electron as it reaches
the grain surface is therefore $E_0+D+Ze^2/a$. In an inelastic collision with a surface
atom/molecule (whose mass is $M_s$) of the grain, the upper limit of energy transfer
is given by
\begin{equation}
\Delta E_u=\frac{4m_e}{M_s}(E_0+\frac{Ze^2}{a}+D)\ ,
\end{equation}
where $m_e$ is the electron mass. From the phonon theory, the probability of
inelastic collision electron collision $\alpha$, and the averaged energy transfer in
each inelastic collision $\delta E$, are given by \citep{UN80}
\begin{equation}
\alpha\approx\frac{3}{2}\frac{\Delta E_u}{kT_D}\ ,\qquad
\delta E\approx\frac{kT_D}{18}\ ,
\end{equation}
where $T_D$ is the Debye temperature of the lattice and is about $420$K for
graphite. The electron has to experience about $l$ inelastic collisions on average
before being truly absorbed, where $l$ is given by
\begin{equation}
l\delta E>E_0\geq(l-1)\delta E\ .
\end{equation}
Between (say, the $i$th and the $(i+1)$th) inelastic collisions, the electrons undergo
elastic collisions with the grain and has a probability $\beta_i$ to escape. Therefore,
the absorption probability is given by \citep{UN80}
\begin{equation}
P_e=\prod_{i=0}^{l-1}\frac{1-\beta_i}{1-\beta_i+\beta_i/\alpha}\ .
\end{equation}

The remaining task is to calculate $\beta_i$, corresponding to the electron escape
probability when its total energy is $E_i=E_0-i\delta E$. This is made convenient by
considering energy conservation to write the radial energy of the electron as 
\begin{equation}
E_r=E_i+\frac{Ze^2}{r}+\frac{e^2a^3}{2r^2(r^2-a^2)}
-\bigg(\frac{a}{r}\bigg)^2(E_i+\frac{Ze^2}{a}+D)\sin^2\theta\ ,
\end{equation}
where the 2nd and 3rd terms represent the potential energy due to net charge and
electric polarization, and the last term represents the transverse energy obtained
from angular momentum conservation $L^2/2mr^2$, with $\theta$ being the angle
between the velocity vector immediately after the elastic collision (scattering) and the
radial direction. We see that $E_r$ approaches infinity close to the grain surface due
to electric polarization, and approaches $E_i$ at infinity. If $E_r$ becomes negative
at some intermediate radius, it means that the electron can not travel beyond this
radius and has to return to the grain surface for another collision. If $E_r$ is positive
at any $r$, the electron trajectory is open and will escape. Therefore, the purpose is
to find the critical value of $\theta_{\rm cr}$, at which the function $E_r(r)$ touches
the horizontal axis tangentially: $E_r=0$ and $dE_r/dr=0$. The latter condition can
be more conveniently replaced by $d(r^2E_r)/dr=0$.

To proceed, we introduce dimensionless variables $x\equiv r/a$,
$\Delta\equiv Da/e^2$ and $\epsilon_i=E_ia/e^2$, and obtain
\begin{equation}
\epsilon_i+\frac{1}{2x^2(x^2-1)}+\frac{Z}{x}=
\frac{1}{x^2}(\epsilon_i+Z+\Delta)\sin^2\theta_{\rm cr}\ ,\label{eq:stick1}
\end{equation}
\begin{equation}
2x\epsilon_i+Z=\frac{x}{(x^2-1)^2}\ .\label{eq:stick2}
\end{equation}
One can solve for $x$ in equation (\ref{eq:stick2}), which always has one solution
for $x>1$. We note that equation (\ref{eq:stick2}) is equivalent to equation (2.10)
of \citet{DraineSutin87} in the calculation of cross section, which is natural
since electron escape is the inverse process of electron capture. Back substituting
the solution of $x$ to equation (\ref{eq:stick1}) one obtains the critical angle
$\theta_{\rm cr}$.

\begin{figure*}
    \centering
    \subfigure{
    \includegraphics[width=75mm,height=60mm]{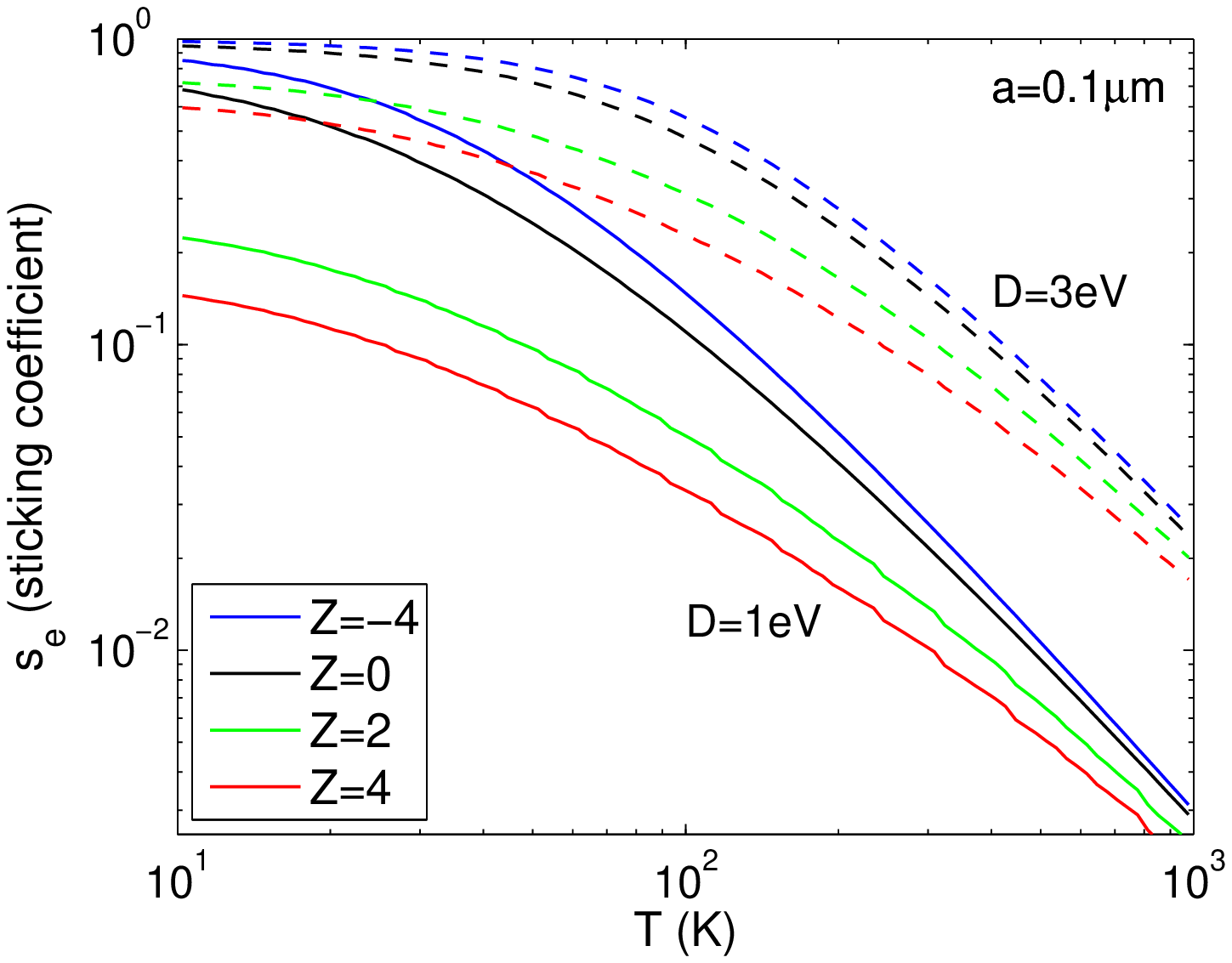}}
    \subfigure{
    \includegraphics[width=75mm,height=60mm]{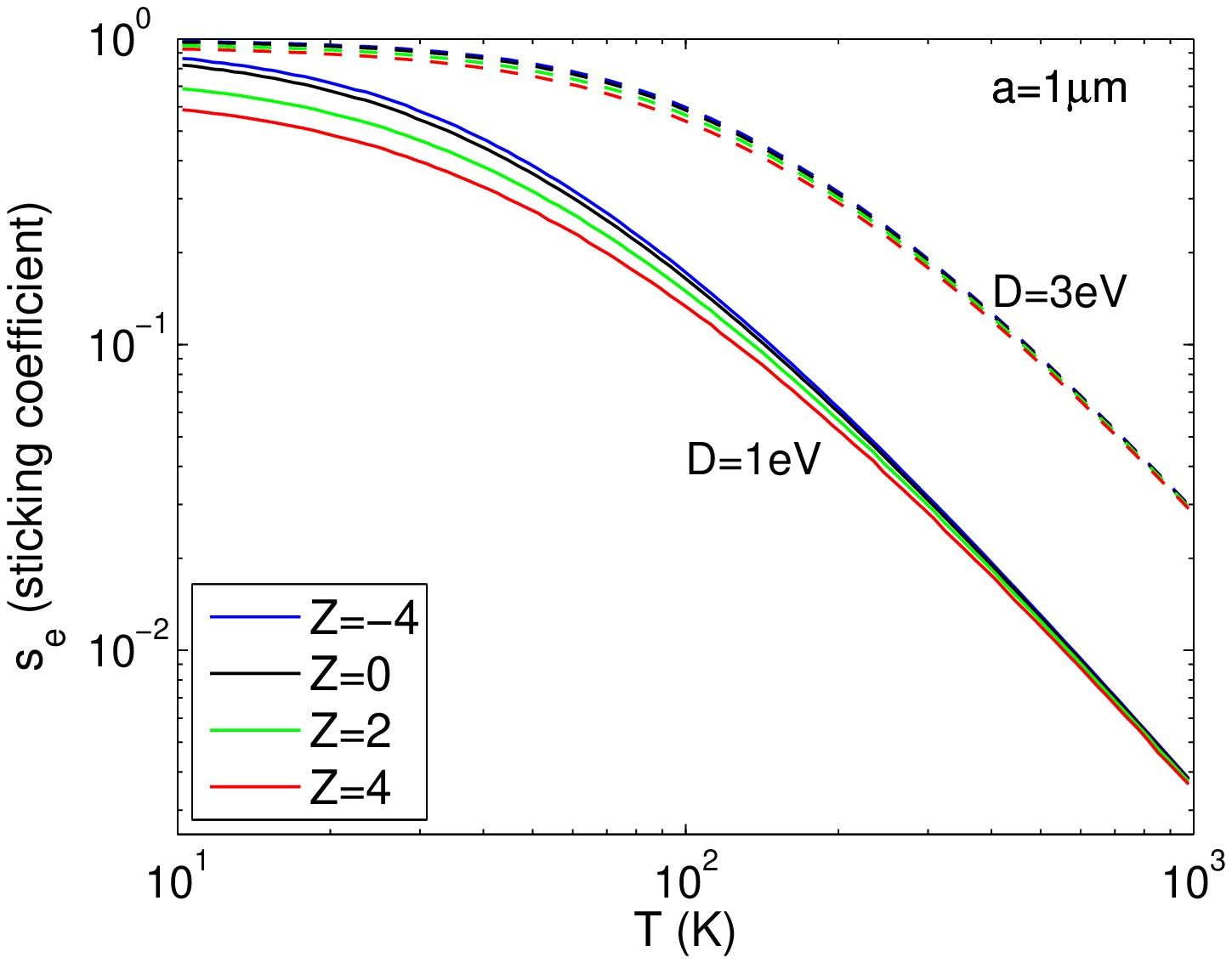}}
  \caption{Electron sticking coefficient for $0.1\mu$m (left) and $1\mu$m (right) sized
  grains  as a function of gas temperature. Two values of the binding energy D=1eV
  (solid lines) and D=3eV (dashed lines) are chosen. In each case, we show curves
  for different grain charges with $Z=-4$ (blue), $Z=0$ (black), $Z=2$ (green), and
  $Z=4$ (red). Their sticking coefficients monotonically decrease as $Z$ becomes
  larger and more positive. We adopt D=1eV for all our chemistry calculations.}\label{fig:estick}
\end{figure*}

One caveat in the calculation is that the left hand side of equation (\ref{eq:stick1})
can become negative when $Z$ is negative and $\epsilon_i$ is small\footnote{
Moreover, the parenthesis on right hand side of equation (\ref{eq:stick1}) can also
become negative. This means that the electron has lost all its kinetic energy before
it reaches the grain surface, and is considered as being ejected.}. In fact, this
expression is proportional to the electron-grain cross section (equation (2.8) of
\citet{DraineSutin87}). This situation correspond to that a potential well with
$E_r<0$ always exist which prevents the electron from approaching the grain (in
the case of electron capture), or escaping from the grain (in the case of we are
interested here). If this occurs, we simply set the escaping probability $\beta_i=0$
(or $\theta_{\rm cr}=0$). In fact, this means that after the $i$th collision, the electron
is destined to be adsorbed even if its total energy $E_i>0$.

Finally, by considering the process of elastic collisions as isotropic scattering, one
obtains the absorption coefficient to be
\begin{equation}
\beta_i=\frac{1}{2\pi}\int_0^{\theta_{\rm cr}}2\pi\sin\theta d\theta
=1-\cos\theta_{\rm cr}\ .
\end{equation}

As an example, we show the electron sticking coefficient for collisions with $0.1\mu$m
and $1\mu$m grains in Figure \ref{fig:estick}. We see that $s_e$ monotonically
decrease with disk temperature and drops by about 2 orders of magnitude from
$10$K to $1000$K. This is because the electron has more excess energy at higher
temperature and has to undergo more inelastic and elastic collisions. Higher
binding energy $D$ raises the sticking coefficient, by reducing the critical angle
$\theta_{\rm cr}$ at each elastic collision. Our new result is that the sticking
coefficient for negatively charged grains is in fact larger than that for positively
charge grains\footnote{This is to be contrasted with Figure 4 of
\citet{IlgnerNelson06}, whose calculation showed that electron sticking is strongly
inhibited when grains are negatively charged. This is incorrect because in their
calculation (using equation (\ref{eq:stickcoeff})), they substituted the original
absorption probability $P_e$ derived in \citet{Nishi_etal91} which is valid only for
collisions with neutral grains. BG09 adopted the same (incorrect) procedure.}. This
is reasonable because for negatively charged grains, the electron has been
decelerated as it approaches the grain, thus bounces fewer times before being
absorbed. Finally, the sticking coefficient is larger for larger grains. Meanwhile,
the difference in the sticking probability for grains with different charged states
is less for larger grains. Both are understandable since the electric potential due to
grain charge at the grain surface reduced for large grains. 

\bibliographystyle{apj}

\label{lastpage}
\end{document}